\documentclass{aa}  
\usepackage{graphicx}
\usepackage{txfonts}

\usepackage{float}
\usepackage{subfloat}
\usepackage{subcaption}
\usepackage[flushleft]{threeparttable}
\usepackage{hyperref}
\hypersetup{colorlinks=true,linkcolor=blue,citecolor=blue,bookmarks=false}

\begin{document}

   \title{SN~2021tsz: A luminous, short photospheric phase Type II supernova in a low-metallicity host}

   \subtitle{}

   \author{R. Dastidar
          \inst{1,2}\fnmsep\thanks{E-mail: rdastidr@gmail.com},
          G. Pignata\inst{2},
          N. Dukiya\inst{3,4},
          K. Misra\inst{3},
          D. A. Howell\inst{5,6},
          M. Singh\inst{7},
          C. P. Guti\'errez\inst{8,9},
          C. Pellegrino\inst{10},
          A. Kumar\inst{11},
          B. Ayala\inst{12,1},
          A. Gangopadhyay\inst{13},
          M. Newsome\inst{14, 5},
          E. Padilla Gonzalez\inst{15},
          K. A. Bostroem\inst{16},
          D. Hiramatsu\inst{17,18,5,6},
          G. Terreran\inst{19},
          C. McCully\inst{5,6}
          }

        \institute{Millennium Institute of Astrophysics (MAS), Nuncio Monsenor Sòtero Sanz 100, Providencia, Santiago RM, Chile
         \and
         Instituto de Alta Investigación, Universidad de Tarapacá, Casilla 7D, Arica, Chile
         \and
         Aryabhatta Research Institute of observational sciencES, Manora Peak, Nainital, 263001, India
         \and
         Department of Applied Physics, Mahatma Jyotiba Phule Rohilkhand University, Bareilly, 243006, India
         \and
         Las Cumbres Observatory, 6740 Cortona Drive Suite 102, Goleta, CA 93117-5575, USA
         \and
         Department of Physics, University of California, Santa Barbara, CA 93106-9530, USA
         \and
         Indian Institute of Astrophysics, Koramangala 2nd Block, Bangalore 560034, India
         \and
         Institut d’Estudis Espacials de Catalunya (IEEC), 08860 Castelldefels (Barcelona), Spain
         \and
         Institute of Space Sciences (ICE-CSIC), Campus UAB, Carrer de Can Magrans, s/n, E-08193 Barcelona, Spain
         \and
         Goddard Space Flight Center, 8800 Greenbelt Rd, Greenbelt, MD 20771, USA
         \and
         Department of Physics, Royal Holloway - University of London, Egham, TW20 0EX, U.K. 
        \and
         Instituto de Astrofísica, Universidad Andres Bello, Fernandez Concha 700, Las Condes, Santiago RM, Chile
         \and
         Oskar Klein Centre, Department of Astronomy, Stockholm University, AlbaNova, SE-106 91 Stockholm, Sweden
         \and
         University of Texas at Austin, 1 University Station C1400, Austin, TX 78712-0259, USA
         \and
         Johns Hopkins University, Bloomberg Center for Physics and Astronomy, San Martin Dr, Baltimore, MD 21210
         \and
         Steward Observatory, University of Arizona, 933 North Cherry Avenue, Tucson, AZ 85721-0065, USA
         \and
         Center for Astrophysics \textbar{} Harvard \& Smithsonian, 60 Garden Street, Cambridge, MA 02138-1516, USA
         \and
         The NSF AI Institute for Artificial Intelligence and Fundamental Interactions, USA
         \and
         Adler Planetarium 1300 S Dusable Lk Shr Dr, Chicago, IL 60605, USA
             }

   \date{Received ; accepted }

 
  \abstract
    {We present the analysis of the luminous Type II Supernova (SN)~2021tsz, which exploded in a low-luminosity galaxy. It reached a peak magnitude of $-$18.88$\pm$0.13\,mag in the $r$ band and exhibited an initial rapid decline of 4.05 ± 0.14 mag (100 d)$^{-1}$ from peak luminosity till $\sim$30 d. The photospheric phase is short, with the SN displaying bluer colours and a weak \ion{H}{$\alpha$} absorption component—features consistent with other luminous, short-photospheric phase Type II SNe. A distinct transition from the photospheric to the radioactive tail phase in the $V$ band—as is common in hydrogen-rich Type II SNe—is not visible in SN~2021tsz, although a modest $\sim$1 mag drop is apparent in the redder filters. Hydrodynamic modelling suggests the luminosity is powered by ejecta-circumstellar material (CSM) interaction during the early phases ($<$30 days). Interaction with 0.6\,M$_\odot$ of dense CSM extending to 3100\,R$_\odot$ reproduces the observed luminosity, with an explosion energy of 1.3$\times$10$^{51}$\,erg. The modelling indicates a pre-SN mass of 9\,M$_\odot$, which includes a hydrogen envelope of 4\,M$_\odot$, and a radius of $\sim$1000\,R$_\odot$. Spectral energy distribution analysis and strong-line diagnostics reveal that the host galaxy of SN~2021tsz is a low-metallicity, dwarf galaxy. The low-metallicity environment and the derived high mass loss from the hydrodynamical modelling strongly support a binary progenitor system for SN~2021tsz. 
  }
  
   \titlerunning{SN 2021tsz}
   \authorrunning{R. Dastidar et al.}
   \maketitle
%
\section{Introduction}

Type II supernovae (SNe) result from the core-collapse (CC) of massive stars that retain their hydrogen envelopes, typically with progenitor masses exceeding 8 M$_\odot$ (e.g., \citealt{Heger2003, Ibeling2013}). They are spectroscopically identified by their prominent hydrogen features. This category includes hydrogen-rich SNe that exhibit either slow or fast declining light curves during the photospheric phase, historically referred to as IIP and IIL, respectively \citep{Barbon1979}. It also includes Type IIb SNe, which initially display hydrogen features before transitioning to helium-dominated spectra \citep{Filippenko1988, Modjaz2019}, and interacting Type IIn SNe, characterised by dominant narrow emission lines arising from interaction of ejecta with circumstellar material (CSM) \citep{Schlegel1990}. Recently, a number of hydrogen-rich Type II SNe have been observed to exhibit `IIn-like' narrow emission features, popularly known as `flash-ionisation' features \citep{Gal-Yam2014}, which, however, last from a few hours to a week \citep{Khazov2016, Bruch2021, Bruch2023, Jacobson2024}, unlike classical Type IIn SNe, where these features last for several weeks. 

Slow-declining hydrogen-rich Type II SNe, traditionally known as the Type IIP SNe, are the most frequently observed, constituting nearly 50\% of all CC-SNe \citep{Li2011, Graur2017, Shivvers2017}. Stellar evolution models suggest red supergiants (RSGs) as the probable progenitors for the slow-decliners \citep{Heger2003, Ekstrom2012, Eldridge2017, Zapartas2019}, with confirmation from progenitor detections in deep pre-explosion images indicating a mass range of 8–18\,M$_\odot$ (\citealt{Smartt2009, Maund2014, Smartt2015, VanDyk2017} and references therein). Among the hydrogen-rich Type II SNe, SNe~1979C \citep{Branch1981, Vaucouleurs1981} and 1980K \citep{Barbon1982A, Buta1982} are regarded as the archetypes of fast-declining Type II SNe. SN~1979C, with a maximum $V$ band magnitude of $-$19.6, is more luminous than most regular SNe II (median magnitude in $V$ band is $-$16.7, \citealt{Anderson2014}), and an analysis of the stellar population surrounding this SN suggests an RSG progenitor with a mass of $\sim$17–18\,M$_\odot$ \citep{VanDyk1999}. On the other hand, for SN~1980K, no star was detected at the SN position in pre-explosion images, possibly indicating a blue supergiant with a mass ranging from 10–15\,M$_\odot$ \citep{Smartt2009} as the likely progenitor. Direct progenitor detection of fast-declining Type II SN has not been as common as for the slow-decliners, and for the few cases that exist, the progenitor identification has been inconclusive (e.g. SN~2009kr, \citealt{Elias2010, Fraser2010, Maund2015}). 

Numerous investigations, including those by \cite{Patat1994} and more recent studies by \cite{Anderson2014, Faran2014, Sanders2015, Valenti2016, Gutirrez2017, Martinez2022, Farfan2024}, have extensively analysed samples of hydrogen-rich Type II SNe. A consistent finding is that faster-declining Type II SNe tend to exhibit higher peak luminosities, shorter photospheric phases, faster expansion velocities, and weaker \ion{H}{$\alpha$} P Cygni absorption compared to their slower-declining counterparts. The latter two characteristics are likely attributable to a lower H-envelope mass \citep{Schlegel1996, Gutirrez2014, Gutirrez2017b} and steeper envelope density gradients in the progenitor \citep{Dessart2005, Dessart2013, Gutirrez2014}, while the higher peak luminosities of fast-declining Type II SNe are suggested to be associated with dense CSM around the progenitor \citep{Blinnikov1993, Morozova2017, Morozova2018, Hillier2019}. 

There is an apparent scarcity of hydrogen-rich Type II SNe with peak absolute magnitudes between $-$18 and $-$20 mag compared to the more common normal-luminosity Type II SNe ($-$16.74$\pm$1.01 mag, \citealt{Anderson2014}). While the peak magnitudes of Type IIn SNe overlap with this range, typically around $-$16.5 to $-$21.5 mag \citep{Li2011, Kiewe2012, Taddia2013, Nyholm2020, Hiramatsu2024}, their luminosity is significantly influenced by ejecta-CSM interaction. Type II SNe that are brighter than $-$20 mag, reaching up to $-$22.5 mag, are classified as superluminous SNe II (SLSNe II; \citealt{Gal-Yam2012, Inserra2018} and references therein). \cite{Arcavi2014} investigated events that bridge the gap between regular Type II SNe and SLSNe II, using the hydrogen-rich core-collapse SN sample from the Palomar Transient Factory (PTF; \citealt{Law2009, Rau2009}). Among the four events identified, only one (PTF10iam) was characterized as a luminous Type II SN with a rapid rise to peak brightness. Later, \cite{Pessi2023} systematically examined the extended Public ESO Spectroscopic Survey of Transient Objects (ePESSTO$+$; \citealt{Smartt2015b}) Marshall from 2017 to 2019 and identified 34 Type II SNe with peak absolute magnitudes within this luminosity gap. Among them, six showed atypical spectral features, lacking the canonical \ion{H}{$\alpha$} P~Cygni profile and instead displaying broad emission with flat absorption. However, their light curves deviated from the norm for Type II SNe, exhibiting no distinct drop from the photospheric phase to the radioactive tail. The prototypical fast-declining Type II SNe~1979C and 1980K, which also show a similar \ion{H}{$\alpha$} profile, do exhibit a distinct transition between the recombination phase and the radioactive tail, unlike the events in the \cite{Pessi2023} sample. A few luminous SNe~II have also been reported to exhibit unusually shorter photospheric phases. \cite{Hiramatsu2021} presented an extensive analysis of three such SNe—2006ai, 2006Y, and 2016egz—with photospheric phases shorter than 70 days. SN~2023ufx and KSP-SN-2022c (a.k.a. AT 2022ozg) are two recent additions to this luminosity gap, which also exhibit a short photospheric phase and spectra lacking \ion{H}{$\alpha$} P Cygni absorption \citep{Tucker2024, Ravi2025, Jiang2025}. 

In regular Type II SNe, the release of shock-deposited energy is believed to be the primary source powering the SN. However, for their luminous counterparts, additional energy sources are necessary. Various powering mechanisms have been suggested for luminous SNe II (e.g. see \citealt{Kasen2017}). Radioactive decay energy from a significant amount of $^{56}$Ni produced during the explosion can significantly contribute to the SN powering mechanism, enhancing observed luminosity. In addition, CSM interaction has been proposed as a potential powering mechanism for several luminous SNe II, sometimes manifested in the early spectra as narrow emission lines persisting for hours to a week. However, the formation of these lines is contingent on CSM density and extent. Some studies have also suggested that some luminous SNe II could be driven by energy deposition from magnetars \citep{Orellana2018}. The luminosity of a SN is also proposed to be correlated with the metallicity of the environment. \cite{Taddia2016} demonstrated that SNe~II with brighter peak magnitudes tend to occur in lower metallicity environments, as massive progenitors are more likely to form in such environments, leading to brighter explosions (but also see \citealt{Gutirrez2018, Grayling2023} where such correlation is not observed).

In this paper, we present a comprehensive analysis and light curve modelling of a luminous SN II, SN~2021tsz, with a fast declining light curve and a short photospheric phase, located in the faint host galaxy SDSS J233758.39-002629.6 (m$_r$\,=\,20.11\,$\pm$\,0.03\,mag, M$_r$\,=\,$-$16.04\,$\pm$\,0.05\,mag). SN~2021tsz (a.k.a ZTF21abmwzxt, ATLAS21bccc, Gaia21dmi, PS21ikp) was discovered by Zwicky Transient Facility (ZTF; \citealt{Bellm2019}) at R.A.(2000) = 23$^{\rm h}$37$^{\rm m}$58$^{\rm s}$.39 and decl.(2000) = $-$00$^\circ$26$^\prime$29$^{\prime\prime}$.38 on 2021 July 19.4 UT (JD 2459414.9) and reported by Automatic Learning for the Rapid Classification of Events (ALeRCE; \citealt{Forster2021}) broker team \citep{Munoz2021}. There is also a close detection from the Asteroid Terrestrial-impact Last Alert System (ATLAS; \citealt{Tonry2018, Smith2020}) survey in the $c$-band on 2021 July 19.5 UT (JD 2459415.0). The SN is located 0.12$^{\prime\prime}$ S, 0.00$^{\prime\prime}$ W from the galaxy centre and is depicted in Fig.~\ref{2021tsz_fits}. ALeRCE's stamp classifier \citep{Carrasco2021} classified the event as an early SN candidate. The discovery magnitude of the event was 18.3 AB mag in the ZTF $g$-filter. The last non-detection with ZTF was on 2021 July 16.5 UT (JD 2459412.0), at a limiting magnitude of 20.2 AB mag in the $r$-band. However, a more recent non-detection was recorded with ATLAS in the $o$-filter on 2021 July 17.6 UT (JD 2459413.1) at a limiting magnitude of 21.0 mag. A spectrum of the SN was obtained by the ePESSTO$+$ team \citep{Smartt2015b} on 2021 August 02.9 UT with the EFOSC2 and Grism 13 (3985-9315~\AA, 18~\AA{} resolution) mounted on the ESO New Technology Telescope (NTT) at La Silla and classified as SN II \citep{Pessi2021}. We will use the mid-point between the last non-detection (from ATLAS) and first detection (from ZTF) as the explosion epoch, which is JD 2459414.0\,$\pm$\,0.9. SN~2021tsz reached a peak absolute magnitude of $-$18.96 in the $g$ band. Noteworthy features of this event include the short photospheric phase, very short fall from the photospheric to the radioactive decay tail and the weak H-absorption component in the spectra during the photospheric phase. 
    
The paper is organised as follows: data reduction procedures are outlined in Section~\ref{sec3}. Section~\ref{sec2} focuses on constraining the explosion epoch and certain host galaxy properties, including distance and reddening. In Section~\ref{sec4}, we delve into the analysis of the light curve, colour curve, and the estimation of the $^{56}$Ni mass ejected in the explosion. Section~\ref{sec:5} is dedicated to spectral analysis, encompassing a comparison of the spectral characteristics with those of other SNe II and literature model spectra. The progenitor, explosion and CSM parameters, including explosion energy, ejecta mass and CSM mass, are determined through bolometric light curve modelling in Section~\ref{sec:6}. In Section~\ref{sec:7}, we compare the parameters from our modelling with those from previous modelling works of similar Type II SNe and discuss the probable progenitor and mass loss scenario. Lastly, the paper concludes with a summary in Section~\ref{sec:8}.

\section{Observations and data reduction}
\label{sec3}

Photometric observations of SN~2021tsz in the optical $BgVri$ bands began approximately 11 days after its discovery and continued till 160 days. These observations were conducted using Sinistro cameras on the 1-m telescopes of the Las Cumbres Observatory (LCO) as part of the Global Supernova Project (GSP) collaboration \citep{Brown2013}. Data collection extended over a period of 160\,days post-discovery. Image pre-processing, including bias correction and flat-fielding, is performed using the BANZAI pipeline \citep{McCully2018}. Subsequent data reduction is handled with the \texttt{lcogtsnpipe} pipeline \citep{Valenti2016}, a photometric reduction tool built on \textsc{Pyraf}. This pipeline calculates zero-points and colour terms and extracts magnitudes using the point-spread function technique \citep{Stetson1987}. Although the SN exploded in a faint host galaxy, its apparent magnitudes during the nebular phase diminished to the luminosity level of the host, making template subtraction necessary to remove host contamination. We performed template subtraction across the entire light curve using host galaxy template images obtained on September 06, 2024, with the PyZOGY image subtraction algorithm \citep{Zackay2016, Guevel2017} integrated within the \texttt{lcogtsnpipe} pipeline. The photometry in the $BV$ bands is presented in Vega magnitudes, while the data for the $gri$ bands are presented in AB magnitudes, calibrated using the American Association of Variable Star Observers (AAVSO) Photometric All-Sky Survey (APASS, \citealt{Henden2016}) catalog. The final calibrated photometric data from LCO are compiled in Table~\ref{Tab1:2021tsz_phot}.

 We also include the ZTF $g$, $r$ and $i$ bands forced photometry data in this study. The ZTF forced-photometry data, obtained from the forced photometry server \citep{Masci2023}, is further processed to discard outliers and large error data points following the procedure in \cite{Hernandez2023}. The ZTF photometry is reported on the AB magnitude scale and is calibrated using Pan-STARRS1 Survey (PS1) sources \citep{Masci2019}. The ZTF data is provided in Table~\ref{Tab2_ZTFphot}. We also obtained photometry from the ATLAS forced-photometry server \citep{Shingles2021} in the ATLAS $c$ and $o$ filters at the position of SN~2021tsz. The fluxes of the observations were averaged using a weighted mean on a nightly cadence and tabulated in Table~\ref{Tab3:atlas_phot}.

Spectroscopic observations of SN~2021tsz were initiated $\sim$7 days after discovery using the low-resolution FLOYDS spectrographs on the 2-m Faulkes Telescope North and South \citep[FTN and FTS,][]{Brown2013} as part of the GSP collaboration. 
The FLOYDS spectra cover a wavelength range of $\sim$3200–10000\,\AA{} with a resolution of $\sim$18 \AA. However, due to fringing effects at the reddest wavelengths, the signal-to-noise ratio beyond 7500~\AA{} is low. The spectra are reduced using the \texttt{floydsspec}\footnote{\href{https://www.authorea.com/users/598/articles/6566}{https://www.authorea.com/users/598/articles/6566}} pipeline, which employs standard reduction techniques. Additionally, the reduced classification spectrum obtained by the ePESSTO+ team is downloaded directly from the TNS (Transient Name Server\footnote{\href{https://www.wis-tns.org/object/2021tsz}{https://www.wis-tns.org/object/2021tsz}}). The reduced spectra are then scaled to the $BgVri$ photometry of the corresponding epoch using the \texttt{lightcurve-fitting}\footnote{\url{https://github.com/griffin-h/lightcurve_fitting}} module \citep{Hosseinzadeh2022}. The log of spectroscopic observations of SN~2021tsz is given in Table~\ref{Tab4:sn_spectra_log}.

\section{Host galaxy properties}
\label{sec2}
\subsection{Distance and extinction}
The host galaxy of SN~2021tsz, SDSS J233758.39-002629.6, is faint with photometric data available from the SDSS archive, but lacks a spectroscopic redshift. To constrain the redshift, we obtained a host galaxy spectrum with the Inamori-Magellan Areal Camera and Spectrograph (IMACS) and grating 150 on the 6.5 m Magellan-Baade telescope \citep{Dressler2011}, with an exposure of 1800s on 2024 July 07. Prominent features include hydrogen Balmer lines, \ion{H}{$\alpha$} and \ion{H}{$\beta$}, and nebular lines such as, [\ion{O}{iii}] $\lambda\lambda$4959,5007 and [\ion{N}{ii}] $\lambda$6583. We estimated the redshift from \ion{H}{$\beta$}, [\ion{O}{iii}] $\lambda\lambda$4959,5007, yielding a mean redshift of 0.0366 $\pm$ 0.0020. As a sanity check, we estimated the redshift from narrow \ion{H}{$\alpha$} emission visible atop the broad \ion{H}{$\alpha$} P-Cygni profile in SN~2021tsz spectra, obtaining 0.0364$\pm$0.0006, where the error is the standard deviation of the measured redshifts. To account for uncertainties in the redshift and the local peculiar motion of the host galaxy, we combine the redshift measurement error with a redshift uncertainty corresponding to a peculiar velocity of 180 km s$^{-1}$, added in quadrature. The peculiar velocity of the galaxy is estimated using the \cite{Carrick2015} peculiar velocity model\footnote{publicly available from http://cosmicflows.iap.fr}. We compute luminosity distances using the \cite{Planck2020} cosmological parameters from the \textsc{astropy.cosmology} package, adopting a flat $\Lambda$CDM cosmology with ${\rm H}_{\rm 0}\,=67.66\pm0.42\,\mathrm{km}\,{{\rm{s}}}^{-1}{\mathrm{Mpc}}^{-1}$ and $\Omega_{\rm m}$ = 0.3111$\pm$0.0056. We propagate these uncertainties using a Monte Carlo approach: we draw 10,000 samples from Gaussian distributions of redshift, ${\rm H}_{\rm 0}$, and $\Omega_{\rm m}$ and compute the luminosity distance for each realization using \textsc{astropy.cosmology}. The final distance and its uncertainty are taken as the median and the 68\% confidence interval of the resulting distribution, which is 166.7 $\pm$ 9.8 Mpc, and the corresponding distance modulus is $\mu$ = 36.10 $\pm$ 0.12 mag.

The Galactic reddening E($\rm{B}-\rm{V}$)$_{\rm MW}$ in the line of sight of the SN is 0.0297$\pm$0.0018 mag, obtained from the extinction dust maps of \cite{Schafly2011}. There is no conspicuous narrow \ion{Na}{ID} absorption in the host galaxy spectrum at the redshift of the host galaxy. Hence, we assume negligible host-galaxy extinction and do not include any contribution to the reddening of the SN from the host.

\subsection{Recent and ongoing star formation rate}
\begin{figure}
\includegraphics[scale=0.27, clip, trim={3.4cm 0.4cm 0cm 2.3cm}]{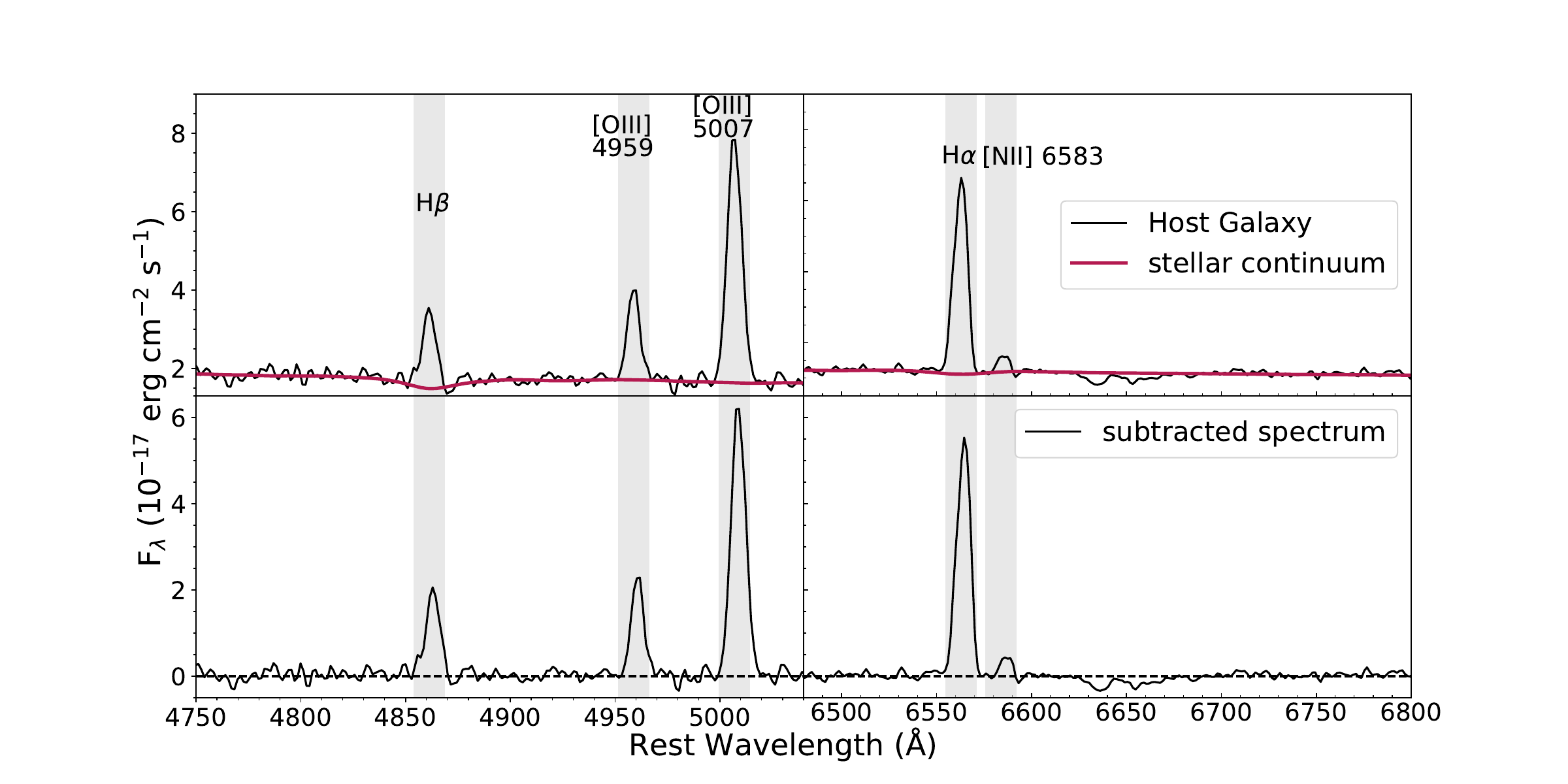}
\caption{Top panel: Spectrum of the host galaxy SDSS J233758.39-002629.6 of SN~2021tsz, overlaid with the best-fit stellar continuum model obtained using \texttt{FIREFLY}. Prominent emission lines, including \ion{H}{$\alpha$}, \ion{H}{$\beta$}, [\ion{O}{iii}] $\lambda\lambda$4959,5007, and [\ion{N}{ii}] $\lambda$6583, are marked. Bottom panel: Host galaxy spectrum after subtraction of the stellar continuum model, highlighting the nebular emission lines used for redshift, star formation rate (SFR), and metallicity diagnostics.}
\label{2021tsz_host}
\end{figure}

To accurately measure emission line fluxes, we fitted the stellar continuum of the host spectrum using \texttt{FIREFLY} (Fitting IteRativEly For Likelihood analYsis, v1.0.3, \citealt{Wilkinson2017, Neumann2022}) with stellar population models from \cite{Maraston2020} and assuming a \cite{Chabrier2003} initial mass function (IMF), and a resolving power of 650. The host spectrum, with the best-fit stellar continuum and subtracted host galaxy spectrum, is shown in the top and bottom panels of Fig.~\ref{2021tsz_host}, respectively. The subtracted host galaxy spectrum was used to measure the emission line fluxes by a Gaussian fit. 

We use the \ion{H}{$\alpha$} and ultraviolet (UV) emission to trace the ongoing (\textless 16 Myr old) and recent (16–100 Myr old) star formation history of the host galaxy, respectively \citep{Gogarten2009}. From the \ion{H}{$\alpha$} flux of 5.2~$\times$~10$^{-16}$~erg~cm$^{-2}$~s$^{-1}$, we derived an ongoing star formation rate (SFR) of 0.0084\,$\pm$\,0.0003\,M$_\odot$\,yr$^{-1}$ using the relationship between \ion{H}{$\alpha$} luminosity and SFR (equation 2) from \cite{Kennicutt1998} and the scaling factor of 0.63 from \cite{Madau2014} to convert from the Salpeter to the Chabrier IMF in the \cite{Kennicutt1998} relation. The far ultraviolet (FUV) magnitude was obtained as 23.31 $\pm$ 0.24 mag from the Galaxy Evolution Explorer (GALEX) source catalog \citep{Seibert2012} accessed through the NASA Extragalactic Database (NED; \citealt{Helou1991}). After correcting for the line-of-sight MW extinction, we transformed the FUV magnitude into SFR using equation 3 of \cite{Karachentsev2013}: 
 $$\log_{10}({\rm SFR} [{\rm M}_\odot {\rm yr}^{-1}]) = 2.78 - 0.4{\rm m}^c_{FUV} + 2\log ({\rm D} [{\rm Mpc}]),$$
 where m$^c_{FUV}$ is the FUV apparent magnitude corrected for total line of sight extinction, and D [Mpc] is the distance to the host galaxy in Mpc.
We obtained $\log_{10}$(SFR [M$_\odot$ yr$^{-1}$]) as $-$2.02 $\pm$ 0.10, corresponding to an SFR of 0.0096\,$\pm$\,0.0022\,M$_\odot$ yr$^{-1}$. Both methods yield comparable SFR values, indicating a stable star formation history over the last 100 Myr. 

\subsection{Spectral energy distribution analysis}
To further assess the physical properties of the host galaxy, we performed Spectral Energy Distribution (SED) analysis using multi-band photometric measurements from SDSS ($ugriz$), GALEX (FUV, NUV), and Wide-field Infrared Survey Explorer (WISE; \citealt{Wright2010}) band 1. The SED was modelled using the \texttt{CIGALE} (Code Investigating GALaxy Emission; \citealt{Boquien2019}) software, assuming a double exponential star formation history, a \cite{Chabrier2003} IMF, the stellar population models of \cite{Bruzual2003}, the \cite{Calzetti2000} attenuation law, and \cite{Dale2014} dust templates. The \texttt{CIGALE} configuration was adopted from Table 9 of \cite{Das2023}. The best fit, chosen using Bayesian inference, yielded a stellar mass of $\log$(M$_\star$/M$_\odot$) = 8.1~$\pm$~0.2, with both recent (100 Myr) and ongoing (10 Myr) SFR of 0.13 $\pm$ 0.08 M$_\odot$ yr$^{-1}$. Consequently, the specific SFR is $\log$(sSFR [yr$^{-1}$])=$-$8.9 $\pm$ 0.3. The value is significantly higher than the SFR derived from \ion{H}{$\alpha$} line flux, however, the latter is known to have limitations as SFR tracers in dwarf galaxies \citep{Bothwell2009}. The SED-derived value falls within the typical sSFR range of dwarf galaxies on the star formation main sequence ($\log$(sSFR) $\sim$ $-$9.6$\pm$0.8 yr$^{-1}$; \citealt{Huang2012}). 

\subsection{Metallicity and nebular line diagnostics}
\label{sec:3.4}
The intensity ratios of flux between [\ion{N}{ii}] $\lambda$6584 and \ion{H}{$\alpha$}, and between [\ion{O}{iii}] $\lambda$5007 and \ion{H}{$\beta$} were found to be $\log$([\ion{N}{ii}]/\ion{H}{$\alpha$}) = $-$1.17 $\pm$ 0.09 and $\log$([\ion{O}{iii}]/\ion{H}{$\beta$}) = 0.52 $\pm$ 0.02, respectively. These ratios are consistent with those of star-forming regions on the BPT diagram \citep{Baldwin1981}. Using these line fluxes, the metallicity of the host galaxy was estimated as 12 + $\log$(O/H) = 8.17 $\pm$ 0.03 dex ($\sim$0.3 Z$_\odot$) using the empirical relationship (Equation 2) from \cite{Marino2013}. The low-metallicity environment of the host galaxy further reinforces the dwarf nature of the host.  The SN and the host galaxy parameters are summarised in Table~\ref{Tab5:SN2021tsz}.

\section{Light curve analysis}
\label{sec4}

\begin{figure}
\includegraphics[width=\columnwidth]{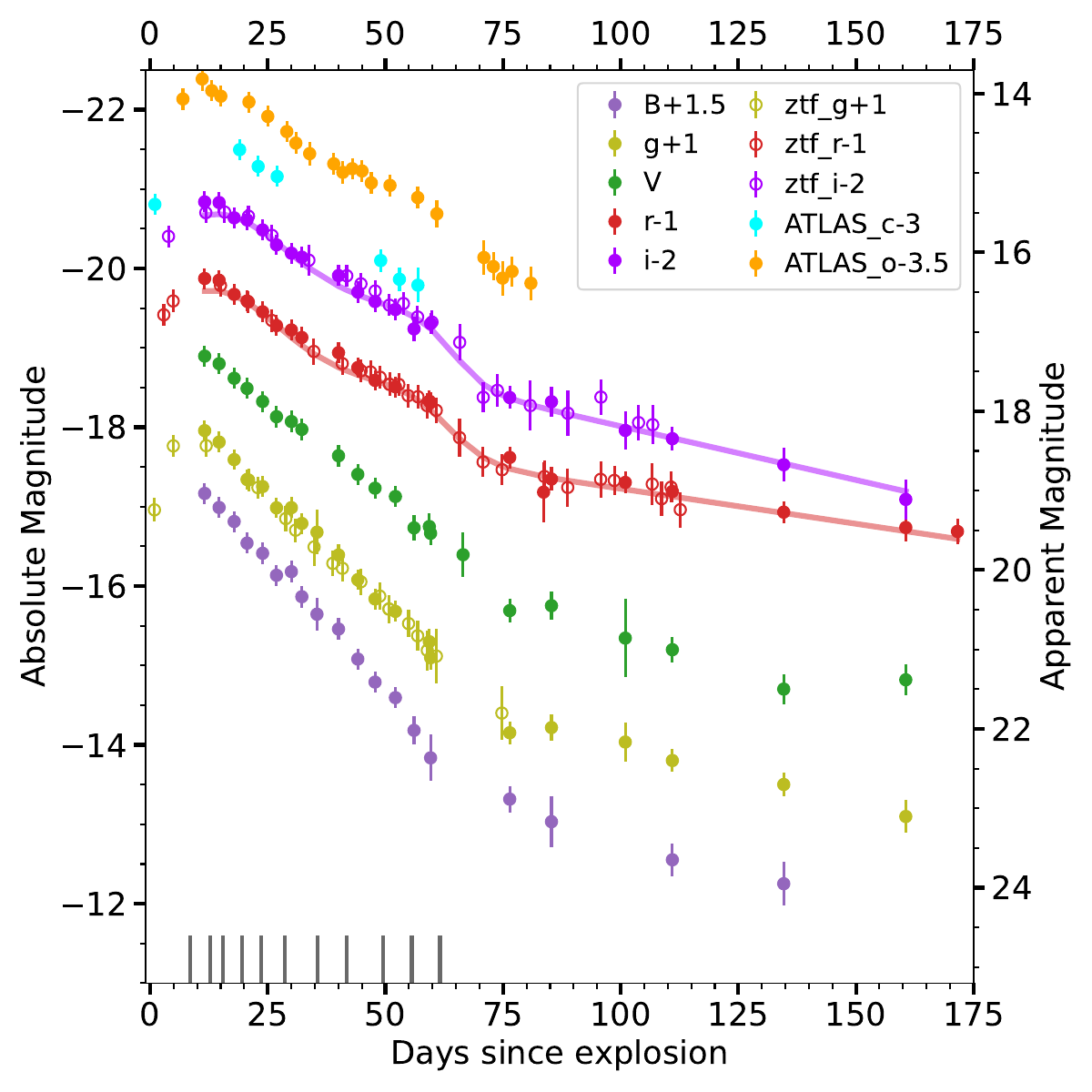}
\caption{$BgVri$, ZTF $gri$, and ATLAS $c$, $o$-band light curves of SN~2021tsz offset by values as shown in the legend. ZTF $gri$ magnitudes are in the SDSS photometric system. The absolute magnitudes are corrected for distance and reddening as listed in Table~\ref{Tab5:SN2021tsz}. Parametrised fit to the $r$ and $i$ band light curves are also shown \citep{Olivares2010}. The vertical gray lines denote the epochs at which a spectrum of SN~2021tsz was obtained.}
\label{2021tsz_LC}
\end{figure}

The $BgVri$ light curves of SN~2021tsz, along with ZTF-$g$, $r$, $i$, and ATLAS $c$ and $o$ bands, are shown in Fig.~\ref{2021tsz_LC}. The light curves peak at absolute magnitudes of $-$18.96$\pm$0.13, $-$18.88$\pm$0.13, and $-$18.85$\pm$0.13 mag in $g$, $r$, and $i$ bands, at 11.6$\pm$0.9, 11.9$\pm$1.6, and 12.4$\pm$1.7 days post-explosion, respectively. \cite{Gonzalez2015} found that Type II SNe rise times span a 1$\sigma$ range of 4–17 days, independent of the filter, placing SN~2021tsz within this distribution. Moreover, \cite{Gall2015} reported a mean rise time of 13.3$\pm$0.6 days for fast-declining SNe II in a sample of 20 core-collapse SNe (CCSNe), which is consistent with the rise time of SN~2021tsz.

After peak brightness, the light curves decline rapidly until $\sim$60 days. Type II SNe decay rates are often characterised by two slopes: an initial steep slope (s$_1$) following peak brightness and a shallower slope (s$_2$), preceding the end of the photospheric phase \citep{Anderson2014}. A break in the $r$-band and ATLAS $o$-band slopes is apparent at 40 days, while other bands show no significant change. A second break, around 50-55 days, is observed across all bands, marking the end of the photospheric phase. This transition from the photospheric phase to the radioactive tail phase is marked by a drop in magnitude, referred to as the transition phase. The transition phase is also marked by a shift towards bluer colours as discussed in Sect.~\ref{Sect4.1}.

We fit the $r$ and $i$ band light curves using Equation 4 from \cite{Olivares2010} to estimate parameters of the transition phase: the magnitude drop ($\mathrm{a_0}$) and the time from the explosion to the transition point between the end of the plateau and the start of the radioactive tail phase, $\mathrm{t_{PT}}$. A distinct drop of ($\mathrm{a_0}$) 0.90$\pm$0.08 mag (0.87$\pm$0.15 mag) occurs between 55-70 days, with $\mathrm{t_{PT}}$ = 63.2\,$\pm$\,1.6\,d (64.6\,$\pm$\,2.2\,d) in the $r$ ($i$) band. The fits are overlaid on the observed $r$ and $i$ band data in Fig.~\ref{2021tsz_LC}. After day 75, the light curves in the $B$, $g$, $V$, $r$ and $i$ bands decline at rates of 1.5, 1.5, 1.4, 0.9, and 1.6 mag (100d)$^{-1}$, respectively. We report the $BgVri$ light curve slopes at different phases in Table~\ref{Tab:LC_slopes}.

To analyse the photometric and spectroscopic properties of SN~2021tsz, we constructed a comparison sample of Type II SNe with peak magnitudes brighter than $-$18 mag and post-peak decline rates exceeding 1 mag (100d)$^{-1}$. This sample comprises the fast-declining prototypes SNe~1979C, 1980K, and 2014G. Additionally, Type II SNe from the literature with photospheric phase durations on the shorter end of the Type II distribution (< 80 days): SNe~2006Y, 2006ai, 2016egz, 2017ahn, 2018gj, 2018hfm, 2020jfo, 2023ufx, and 2024bch, are included irrespective of their peak luminosity. Table\,\ref{comp_sample} lists the comparison sample along with their parameters and corresponding references.

\begin{figure}
\centering
\vspace*{-0.cm}
\includegraphics[width=0.5\textwidth, clip, trim={0.4cm 0.4cm 0cm 0.2cm}]{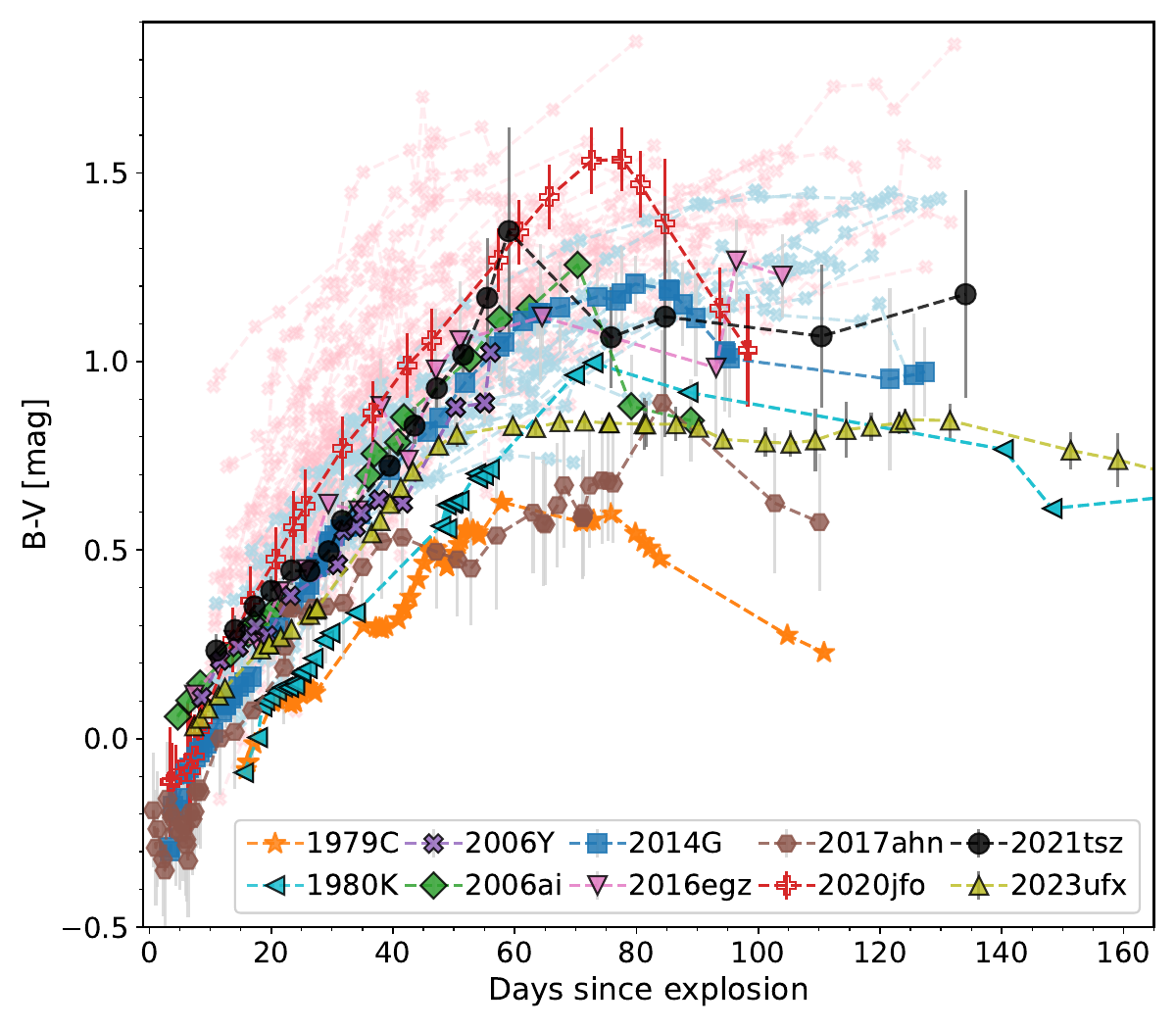}

\caption{$B-V$ colour curve of SN~2021tsz compared to those of the comparison sample SNe. The colour of SN~2021tsz and the comparison sample are extinction-corrected. Prior to estimating the colour, LOESS (locally estimated scatterplot smoothing) regression was applied to smooth out the $B$ and $V$ band light curves of all SNe. In the background, the unreddened sub-sample of 19 SNe II from \citet{dejaeger2018a} is shown in blue, while the rest of the sample SNe with extinction correction applied are shown in pink.}
\label{2021tsz_Color}
\end{figure}

\subsection{Colour evolution}
\label{Sect4.1}
The evolution of extinction-corrected $B-V$ colour for SN~2021tsz is shown in Fig.~\ref{2021tsz_Color}, alongside SNe from the comparison sample and the regular SN II sample from \cite{dejaeger2018a} (hereafter J18). The J18 sample consists of an unreddened sample of 19 SNe II with negligible host extinction, while the rest are extinction-corrected colours of SNe II. A non-parametric smoother, LOESS-locally weighted running line smoother \citep{Cleveland1979}, was applied to fit the $B$ and $V$ band light curves before estimating the colours for all comparison SNe. The $\mathrm{B-V}$ colour of SN~2021tsz gradually shifts to redder colours, from 0.23 to 1.35 mag, until $\sim$ 60 days post-explosion. This evolution is consistent with the trend observed in the J18 sample. The SNe in the comparison sample, while following a similar redward evolution, generally exhibit comparatively bluer colours.

\begin{figure}
    \centering
        \includegraphics[width=0.55\textwidth, clip, trim={0cm, 1.2cm, 0cm, 2.85cm}]{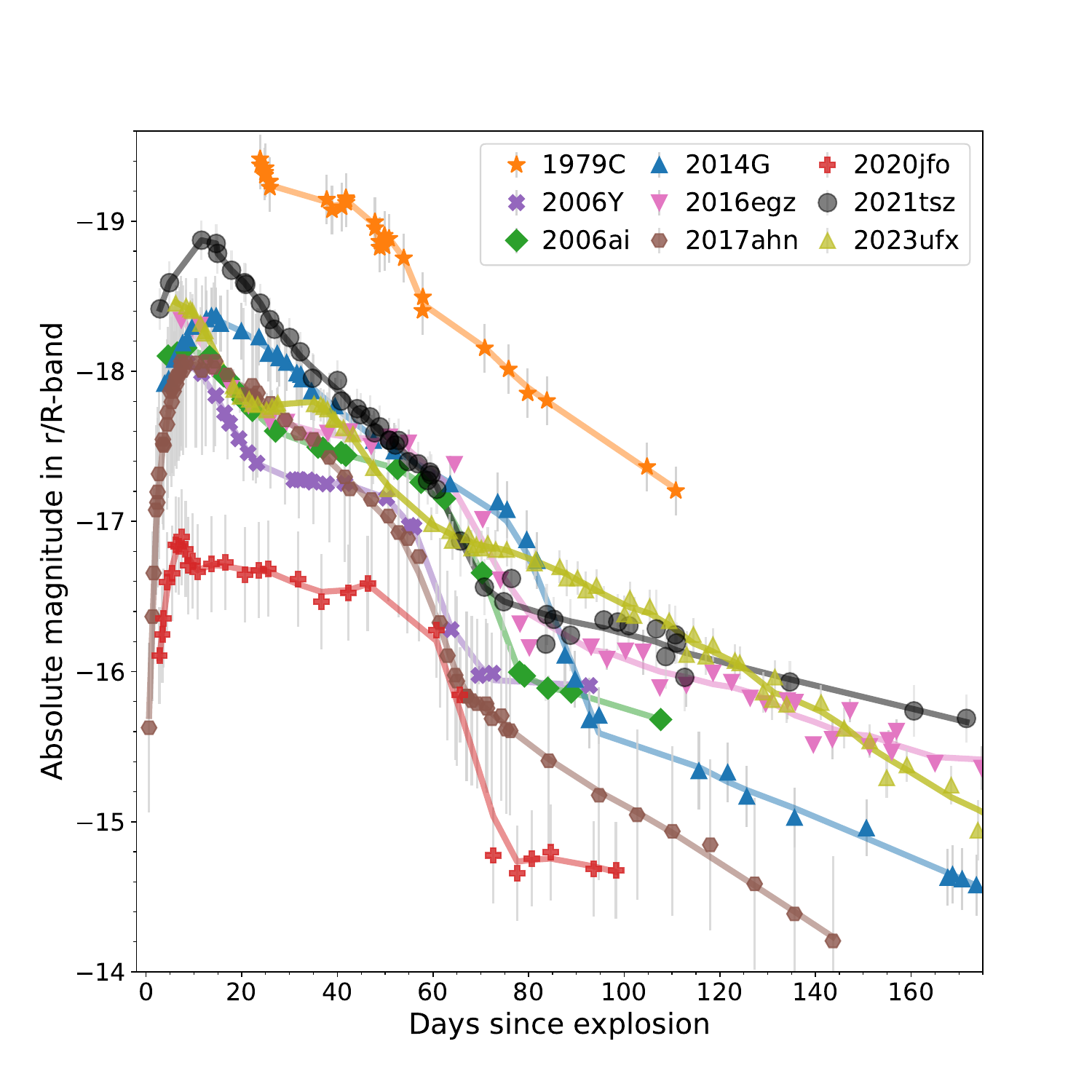}
        \label{fig:figure2}
    \caption{Comparison of absolute $r$ band light curve of SN~2021tsz with those of the comparison sample. The magnitudes are corrected for distance and reddening as listed in Table~\ref{comp_sample}.}
    \label{2021tsz_Absmag}
\end{figure}

Between 60 and 80 days, SN~2021tsz undergoes a transition towards bluer colours, shifting from 1.35 to 1.10 mag. Subsequently, it shows little change, maintaining a nearly constant colour of 1.1 mag. This transition toward bluer colours is characteristic of all SNe II as they enter the nebular phase. At this phase, the optically thin envelope can no longer thermalise the photons from radioactive decay, resulting in bluer colours. The timing of this transition coincides with the light curve evolution from the photospheric phase to the radioactive tail phase, occurring around 70 days in SN~1980K, 85 days in SN~2014G, and 60 days in SN~2021tsz (also see Fig.~\ref{2021tsz_Absmag}). After 80 days, the $B-V$ colour of SN~2021tsz is similar to SN~2014G, while most other SNe in the comparison sample exhibit much bluer colours. Overall, while luminous SNe II tend to exhibit bluer $B-V$ colours, their colour evolution is similar to that of traditional SNe II. The bluer colours may be attributed to interactions with the surrounding CSM, which generates additional thermal energy. Finally, we note that CSM interaction can also result in a bluer colour during the nebular phase. However, the blueward turn associated with CSM interaction is expected to be slow and gradual, whereas the change observed here is more rapid and coincides with the expected timing of the end of recombination phase.

\subsection{Absolute magnitude and \texorpdfstring{$^{56}$Ni}{Lg} mass} \label{sect:4.2}

The evolution of the $R/r$ band absolute magnitude for SN~2021tsz is shown alongside SNe from the comparison sample in Fig.~\ref{2021tsz_Absmag}. The peak $r$-band magnitude of SN~2021tsz is brighter than most SNe in the comparison sample, with the exception of SN~1979C. Some events, such as SNe~2017ahn and 2021tsz, exhibit a linearly declining light curve during the photospheric phase, while others—like SNe~2006Y and 2023ufx—show a rapidly cooling phase followed by a plateau. The short photospheric phase sample spans a wide magnitude range, with SN~2020jfo at the faint end, being $\sim$2 mag fainter than SN~2021tsz, and SN~1979C at the bright end. Compared to the well-observed, fast-declining SN~2014G, SN~2021tsz is $\sim$0.5 mag brighter at peak, has a shorter photospheric phase and a smaller luminosity drop at the end of this phase than SN~2014G.

The amount of $^{56}$Ni synthesised during an explosion is typically estimated from the luminosity of the radioactive tail phase in most Type II SNe. We use the bolometric correction (BC) method to derive the tail bolometric luminosity and estimate the $^{56}$Ni mass. While \cite{Hamuy2003} provided a BC for the $V$ band based on three SNe II, more recently, \cite{Rodriguez2021} analysed a sample of 15 SNe II with $BV(r)R(i)IJHK$ photometry and three theoretical spectral models to calibrate the BC values. They concluded that the $i$ band is better suited for estimating $^{56}$Ni mass compared to the $V$ and $r$ bands, as the constant BCs in the latter overestimate the tail bolometric luminosity and, consequently, the $^{56}$Ni mass for moderately luminous SNe II. Using Sloan {\it i} band magnitudes from 101 to 134.6 days post-explosion and the BC values provided in Table 9 of \cite{Rodriguez2021}, we estimated a $^{56}$Ni mass of 0.08$\pm$0.01\,M$_\odot$. This coincides with the upper end of the distribution of $^{56}$Ni masses for Type II SNe \citep{Muller2017, Anderson2019, Rodriguez2021}. For comparison, $^{56}$Ni masses derived from $V$ and $r$-band tail luminosities in the same time range are 0.10$\pm$0.02 and 0.13$\pm$0.02 M$_\odot$, respectively, both higher than the estimation from the $i$-band. Therefore, for light curve modelling in Sect.~\ref{sec:6}, we adopt the $^{56}$Ni mass estimated from the $i$-band magnitudes. We note, however, that in cases of strong CSM interaction during the nebular phase, this method may yield an overestimate of the actual $^{56}$Ni synthesised in the explosion. The value we report should therefore be regarded as an upper limit.

\section{Spectral analysis} \label{sec:5}
\subsection{Spectral evolution}
\begin{figure*}
    \centering
    \begin{subfigure}[b]{0.5\textwidth}
        \centering
        \includegraphics[scale=0.48, clip, trim={0cm 0 0 0.5cm}]{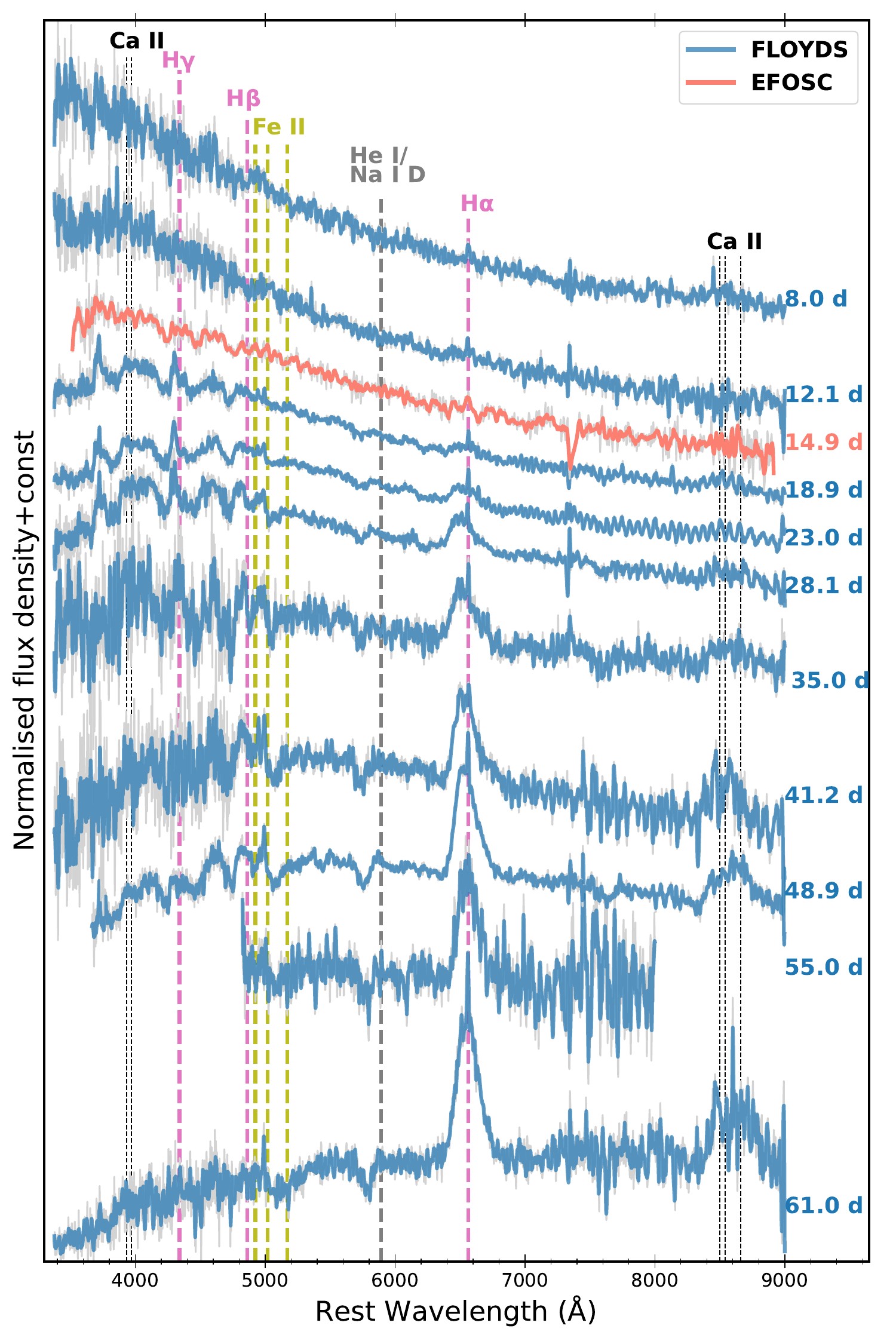}
    \end{subfigure}
    \hfill
    \begin{subfigure}[b]{0.4\textwidth}
        \centering
        \includegraphics[scale=0.35, clip, trim={0cm 0 0.4cm 0.3cm}]{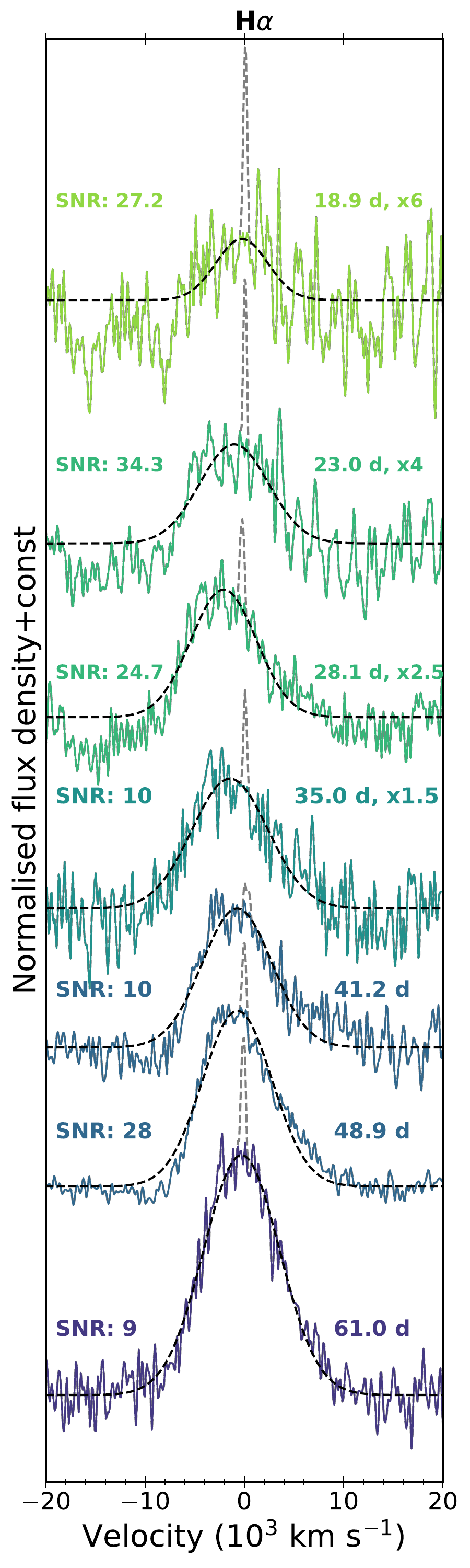}
    \end{subfigure}
    
    \caption{Spectral evolution of SN~2021tsz from 8.0 to 61.0\,days after explosion, with prominent lines marked. Right panel: Evolution of the \ion{H}{$\alpha$} profile in velocity space (centred at the rest wavelength of \ion{H}{$\alpha$}) for spectra with SNR $>$ 5. The broad component is shown after subtraction of the narrow Gaussian (FWHM $\sim$10 \AA) with its best-fit profile overplotted with black dashed lines. The corresponding narrow component, attributed to the host galaxy, is indicated by gray dashed lines.}
    \label{2021tsz_spectra}
\end{figure*}

\begin{figure*}
\centering
\includegraphics[scale=0.5, clip, trim={0 0 0 0cm}]{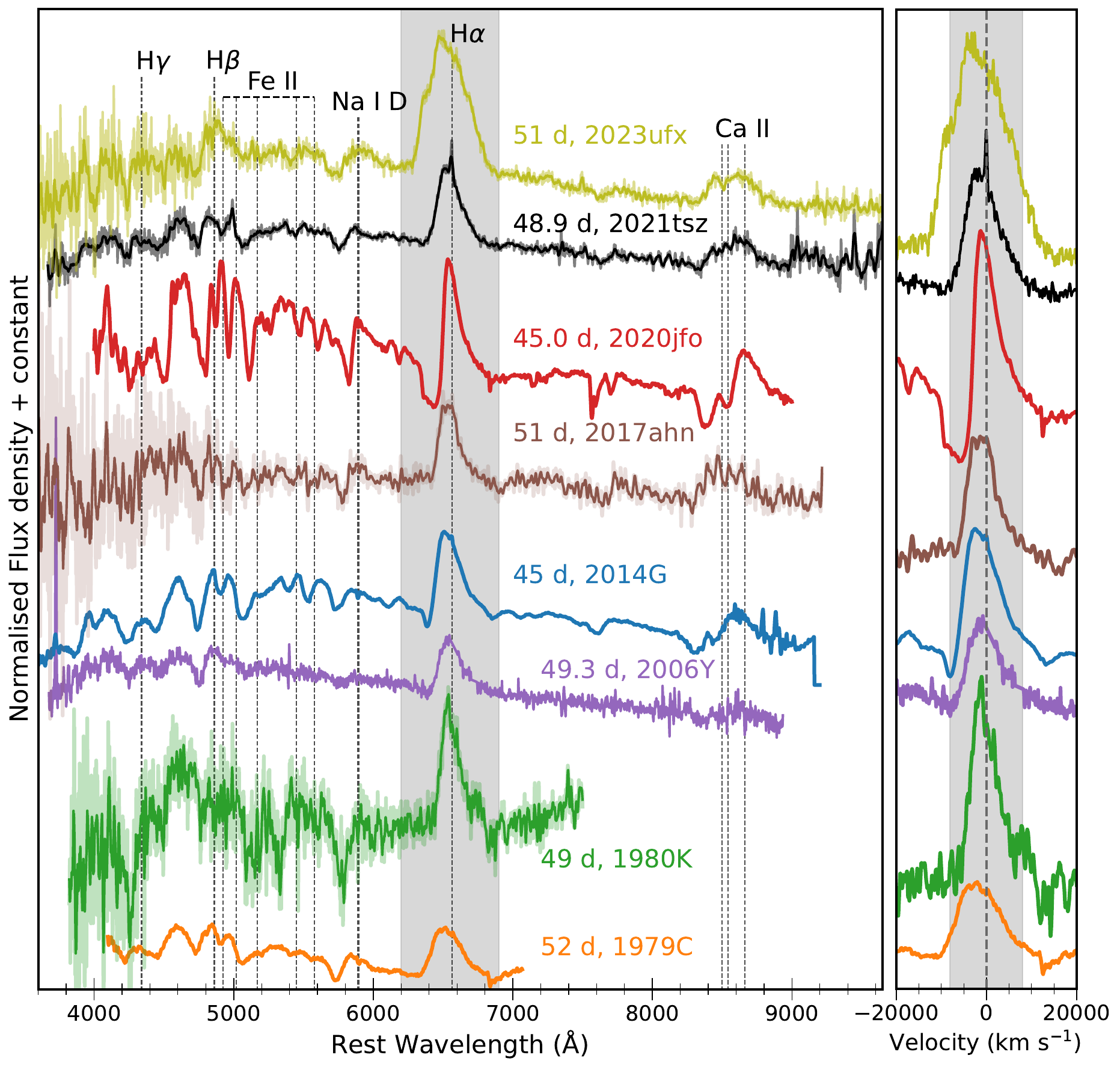}
\caption{Comparison of 48.9\,day spectrum of SN~2021tsz with similar epoch spectra of the comparison sample corrected for both redshift and reddening. On the right, cut out of the \ion{H}{$\alpha$} region is shown in the velocity space, centred at the rest wavelength of \ion{H}{$\alpha$}.}
\label{2021tsz_spectra_comp_50}
\end{figure*}

The spectral evolution of SN~2021tsz from 8.0 to 61.0\,days is shown in Fig.~\ref{2021tsz_spectra}. Up to 14.9 days, the spectra exhibit a blue continuum, with the \ion{H}{$\alpha$} emission line becoming prominent in the 18.9 day spectrum. \ion{H}{$\beta$} and \ion{H}{$\gamma$} absorption lines are also evident at this epoch. A narrow emission atop the broad \ion{H}{$\alpha$} profile is consistently observed at the same wavelength, likely originating from host galaxy \ion{H}{$\alpha$} emission. An absorption feature around 5700\,\AA{} emerges in the 18.9\,day spectrum, which may correspond to \ion{He}{i}, \ion{Na}{id}, or a combination of both. Beyond one month, the spectra are dominated by strong and broad \ion{H}{$\alpha$} emission (full width at half maximum, FWHM velocity, v $\approx$ 8000-9000 km s$^{-1}$), with minimal to no P Cygni absorption. The emergence of \ion{Fe}{ii} $\lambda$5169 and the near infrared (NIR) \ion{Ca}{ii} triplet is evident in the 28.1\,day spectrum. 

At early times, the dense and hot ejecta result in a higher population of hydrogen atoms in the n=2 state (responsible for \ion{H}{$\beta$}), leading to stronger \ion{H}{$\beta$} absorption compared to \ion{H}{$\alpha$} under cooler conditions. However, in SN~2021tsz, the \ion{H}{$\alpha$} absorption component is weak even during the photospheric phase. This may be attributed to photons generated from interaction of CSM with high-velocity outer ejecta filling the \ion{H}{$\alpha$} absorption trough. \ion{H}{$\beta$}, \ion{H}{$\gamma$}, and \ion{H}{$\delta$} are not similarly affected because the CSM is still in its early phase and optically thick. Thus, the high optical depth efficiently converts the higher-order Balmer-series photons from the interaction into \ion{H}{$\alpha$}, resulting in a steep Balmer decrement \citep{Chevalier1994, Zhang2022}.

\citet{Dessart2022} suggested that a spherical dense shell swept up by the shock may generate high-velocity emission that fills the blue-shifted absorption and produces a red excess in emission. This effect depends on the density of the shell, which sets the shock power and, in turn, the strength of these features. In SN~2021tsz, the absence of a red excess and a weak \ion{H}{$\alpha$} absorption component may instead indicate an asymmetric shock-swept dense shell, pointing to an aspherical shock and/or an aspherical CSM configuration.

In the right panel of Fig.~\ref{2021tsz_spectra}, the evolution of \ion{H}{$\alpha$} in the velocity space is shown for the spectra with signal-to-noise ratio (SNR) greater than 5. The profile is initially fitted with two Gaussian components, a narrow and a broad one. The narrow component, consistently well-reproduced with a Gaussian of FWHM $\sim$10 $\AA$ across all epochs, most likely originates from the host galaxy. In the figure, we present the broad component after subtracting this narrow contribution; the fit to the broad profile is overplotted, while the narrow component atop the broad profile is shown as gray dashed lines. The broad \ion{H}{$\alpha$} peak is blue-shifted until 48.9 days, after which it becomes more symmetric about the rest wavelength. Such early blue-shifts are typical in Type II SNe, arising from resonance photon scattering in an atmosphere dominated by electron scattering opacity \citep{Chugai1988, Jeffery1990}. Alternatively, they have been linked to steep ejecta density profiles that enhance occultation of the receding material \citep{Anderson2014b}, with spectral models confirming that such profiles naturally reproduce the observed blueshifted lines \citep{Dessart2005}. 

\subsection{Comparison with other SNe}

In Fig.~\ref{2021tsz_spectra_comp_50}, we compare the photospheric spectrum of SN~2021tsz at 48.9 day with those of the comparison sample SNe at similar epochs. All the spectra are de-redshifted and dereddened before comparison. The spectrum of SN~2021tsz closely resembles those of SNe~1979C, 2006Y, and 2017ahn. The spectrum of SN~2021tsz lacks prominent \ion{H}{$\alpha$} absorption and displays weak \ion{Fe}{ii} lines, similar to the comparison sample, except for SNe~2014G and 2020jfo. The suppression of \ion{H}{$\alpha$} absorption typically suggest the presence of a massive, dense shell formed by swept-up material \citep{Dessart2016}. Additionally, the weaker \ion{Ca}{ii} NIR absorption in SN~2021tsz as well as in the other short photospheric phase Type II SNe~2006Y, 2017ahn and 2023ufx compared to SNe~2014G and 2020jfo indicates that the interaction power injected into the outer ejecta layers increased the abundance of Ca$^{2+}$, leading to the quenching of the \ion{Ca}{ii} NIR absorption \citep{Dessart2022}. A stronger interaction power can cause complete ionisation, resulting in very weak or absent \ion{Ca}{ii} features, as observed in SN~2006Y. The spectrum of SN~2021tsz also shows fewer lines below 6000~\AA{} compared to SN~2014G. This could imply either a higher temperature for SN~2021tsz relative to SN~2014G or a metallicity effect \citep{Dessart2014, Anderson2016, Gutirrez2018}. Since the temperature estimated from the continuum of the 45\,day spectrum of SN~2014G ($\sim$5200\,K) is comparable to that of SN~2021tsz, the latter explanation -- metallicity -- is more plausible. This interpretation is consistent with the low-metallicity of SN~2021tsz's host. Overall, the spectrum of SN~2021tsz is akin to other SNe II with fast-declining light curves and short photospheric phases, e.g., SNe~2006Y, 2006ai, and 2016egz. However, it is markedly different from the slow-declining, short-photospheric-phase SN~2020jfo, which shows strong absorption of \ion{H}{i} and other metal lines. The cut-out on the right, which focuses on the \ion{H}{$\alpha$} region in velocity space, highlights the area corresponding to twice the FWHM velocity of \ion{H}{$\alpha$} emission in SN~2021tsz. With the exception of SN~2023ufx, whose spectrum exhibits a distinct blue excess in the \ion{H}{$\alpha$} emission, neither SN~2021tsz nor the comparison sample shows such excess in the 50 day spectrum. 

\begin{figure}
\centering
\includegraphics[scale=0.4, clip, trim={0.8cm 1.0cm 0.25cm 1.0cm}]{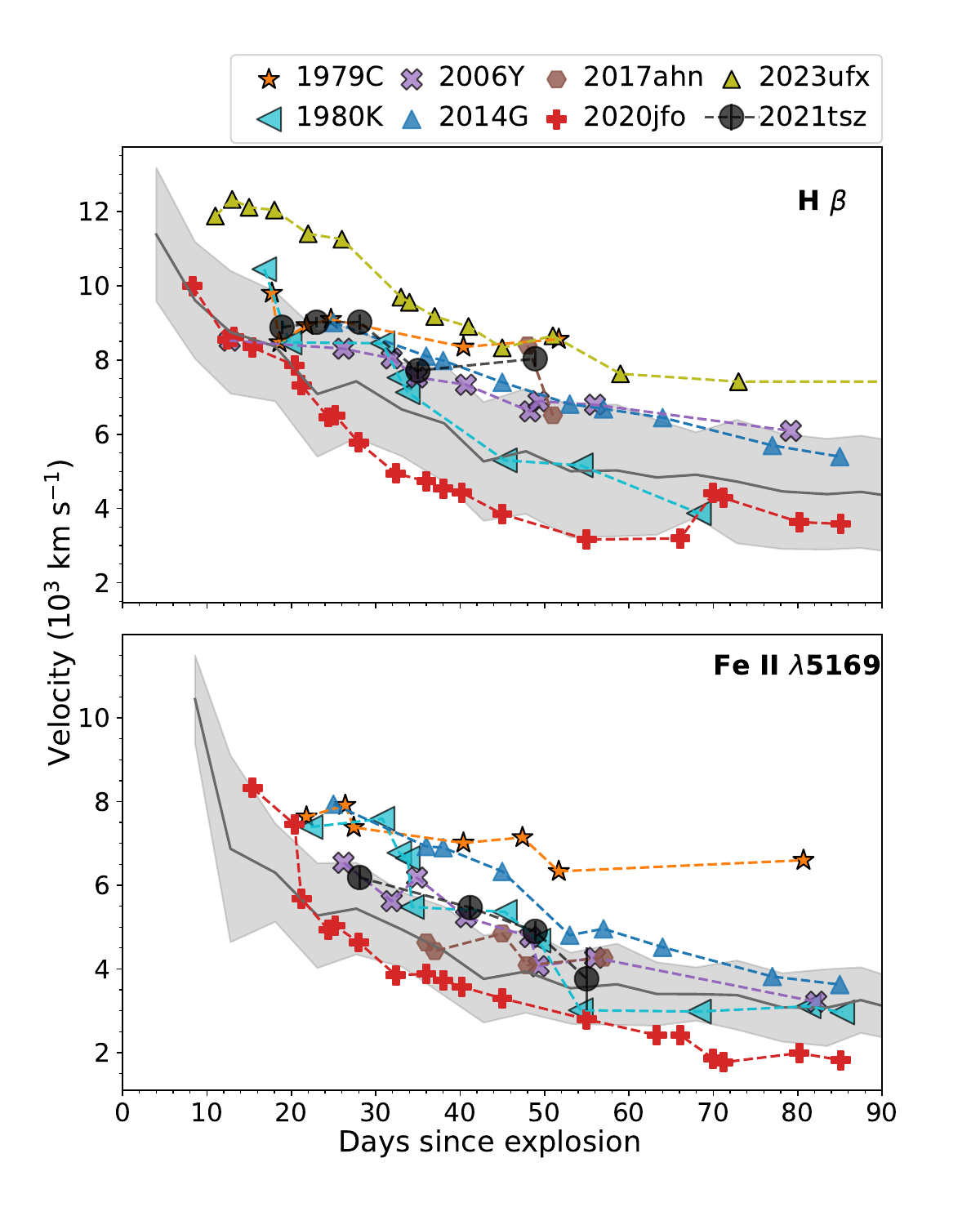}
\caption{Expansion velocity of SN~2021tsz is shown along with that of the comparison sample SNe and the mean velocity (grey) of 122 IIP/IIL SNe \citep{Gutirrez2017} for \ion{H}{$\beta$} (top panel) and \ion{Fe}{ii} $\lambda$5169 (bottom panel). The light grey regions represent the standard deviations of the mean velocities of the sample.}
\label{2021tsz_velocity_comparison}
\end{figure}

We compare the velocity evolution of \ion{H}{$\beta$} and \ion{Fe}{ii} $\lambda$5169, determined from their respective blue-shifted absorption minima, in the spectra of SN~2021tsz with those SNe from our comparison sample and the mean velocities from the Type II sample in \cite{Gutirrez2017} in Fig.~\ref{2021tsz_velocity_comparison}. The \ion{H}{$\beta$} velocities of SN~2021tsz are at the upper end of the velocity distribution of the \cite{Gutirrez2017} sample, similar to SNe~2006Y and 2014G, but lower than SN~2023ufx. Similarly, the \ion{Fe}{ii} $\lambda$5169 velocity of SN~2021tsz lies at the upper end of the corresponding distribution from \cite{Gutirrez2017}, comparable to SN~2006Y but lower than SN~2014G at the early phases. Furthermore, we note that SN~2021tsz and other fast-declining, short photospheric phase SNe~II have higher velocities than the slow-declining, short photospheric phase SN~2020jfo.

\section{Hydrodynamical modelling of lightcurve} \label{sec:6}

SN~2021tsz displays a high peak luminosity, a steep decline after maximum brightness, and a short photospheric phase. To investigate the origin of these characteristics and gain insight into the progenitor and explosion properties, we carried out hydrodynamical modelling of the light curve using the open-source 1D radiation hydrodynamics code, Supernova Explosion Code (\texttt{SNEC}, \citealt{Morozova2015}). \texttt{SNEC} operates under local thermodynamic equilibrium (LTE) conditions and applies grey opacities, without generating spectral calculations. The code requires input parameters such as a progenitor model, explosion energy, $^{56}$Ni mass, and the degree of $^{56}$Ni mixing to produce outputs that include bolometric and multi-band light curves, as well as the evolution of photospheric velocity and temperature. To explore the influence of CSM on the light curves, CSM with varying densities and radial distributions around the progenitor can be added. This setup enables simulations of explosions within CSM environments. 

\begin{table*}
\centering
\caption{Pre-supernova structure and CSM parameters for two progenitor models.} \label{comp_model}
\renewcommand{\arraystretch}{1.2}
\footnotesize
\begin{tabular}{lccccccccccc}
\hline
Progenitor Model  & M$_{\rm pre-SN}$ & R$_\star$  & M$_{\rm H}$ & M$_{\rm He}$ & t$_{\rm offset}$ & E$_{\rm exp}$ & M$_{\rm Ni}$ & K$_{\rm CSM}$ & R$_{\rm CSM}$ & M$_{\rm CSM}$ & $\chi^2_{\rm min}$ \\
                     & (M$_\odot$) & (R$_\odot$) & (M$_\odot$) & (M$_\odot$)  & & (days) & ($10^{51}$ erg) & (M$_\odot$) & ($10^{17}$ g cm$^{-1}$) & (R$_\odot$) & (M$_\odot$) \\
\hline
stripped\_2M & 10.25 & 1013 & 5.16 & 5.09 & $-$0.9 & 1.9 & 0.07 & 4 & 3100  & 0.4 & 14.1 \\
stripped\_3M  & 9.17   & 991 & 4.06 & 5.11 & $-$0.9 & 1.3 & 0.07 & 7 & 3100  & 0.6 & 12.3\\
\hline
\end{tabular}
\end{table*}

Massive stars lose significant amounts of mass during their lifetimes through various mechanisms such as line-driven winds, stable or unstable binary mass transfer, and pulsational instabilities or eruptions (e.g., \citealt{Smith2014}). To capture the effects of this mass loss on the pre-SN structure and the resulting light curve, \citet{Morozova2015} used the stellar evolution code \texttt{MESA} (\citealt{Paxton2019, Jermyn2023}, and references therein) to generate a suite of pre-supernova models for a 15\,M$_\odot$ zero-age main sequence (ZAMS) star, stripped to varying degrees during the middle of the subgiant branch (mSGB) phase. This allows for the exploration of a range of hydrogen envelope masses and corresponding evolutionary outcomes. Previous hydrodynamical modelling of short-photospheric phase Type II SNe suggests that their progenitors retain significantly less hydrogen envelope mass than typical Type IIP SNe \citep{Eldridge2018, Hiramatsu2021, Ravi2025}. Therefore, given the fast declining short photospheric phase of SN~2021tsz, we adopt the stripped-mSGB series from \citet{Morozova2015} for light curve modelling.

\begin{figure}
\begin{minipage}[h]{1.0\linewidth}
\centering
\includegraphics[scale=0.43, clip, trim={0.3cm 0.cm 0.3cm 1.cm}]{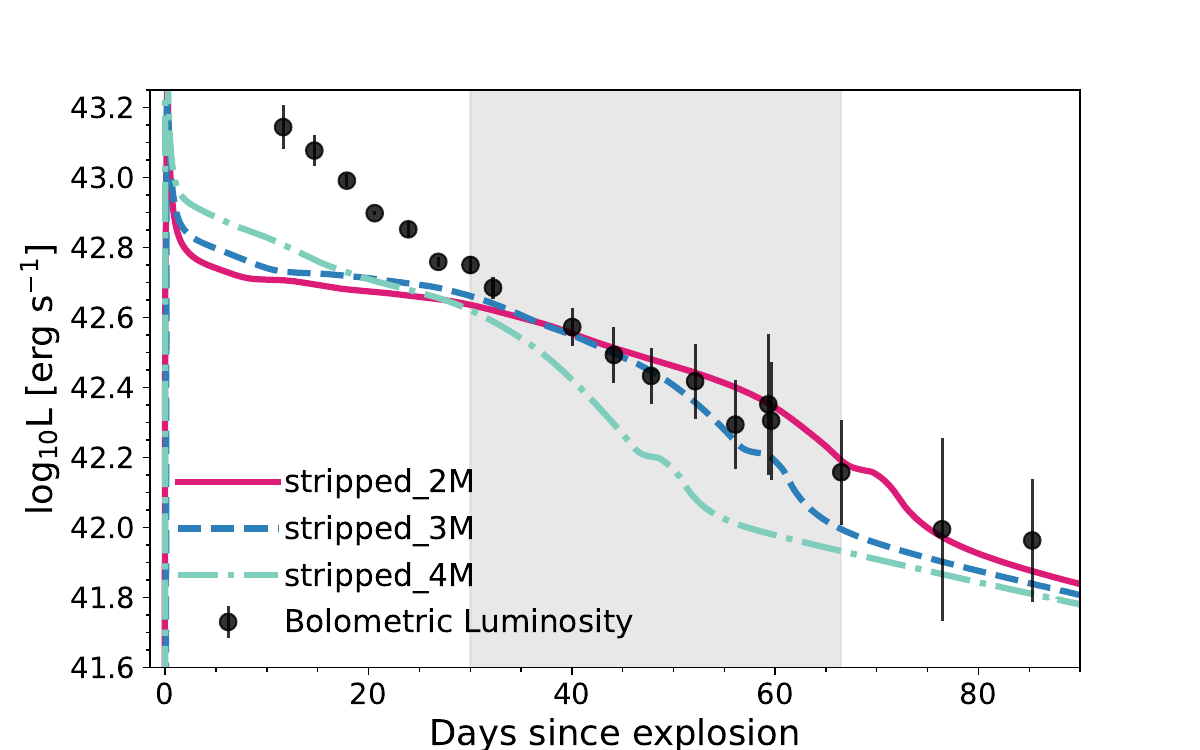}
\end{minipage}

\begin{minipage}[h]{1.0\linewidth}
\centering
\includegraphics[scale=0.45, clip, trim={0.4cm 0.cm 0.3cm 0.cm}]{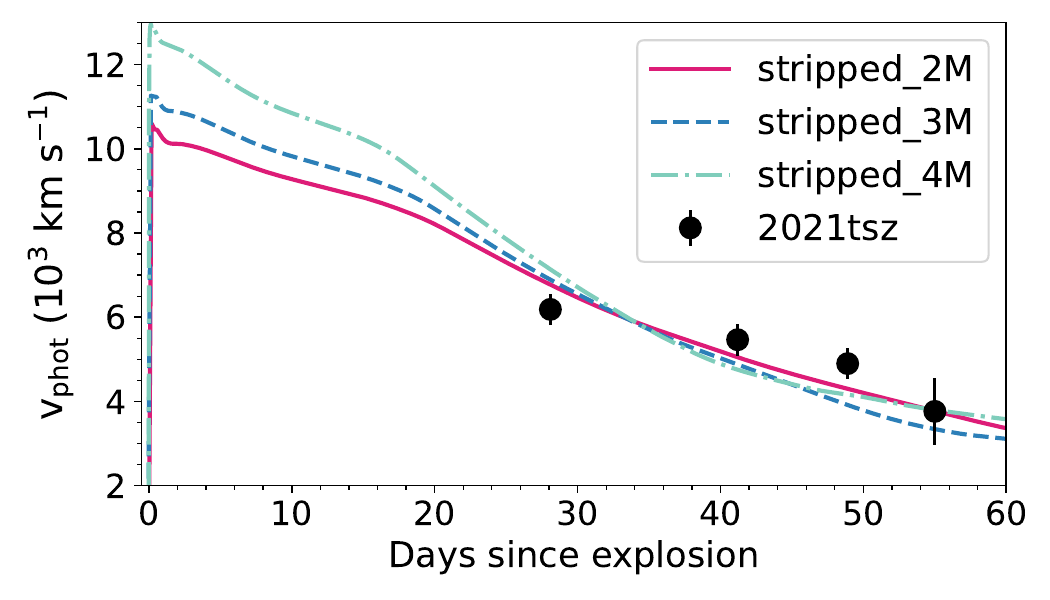}
\end{minipage}
\caption{Bolometric light curve models simulated with \texttt{SNEC} without CSM for 2, 3, and 4 M$_\odot$ stripped masses and fixed explosion energy and $^{56}$Ni masses of 1.3$\times$10$^{51}$ erg and 0.08 M$_\odot$, respectively, are shown along with the observed light curves in the top panel. The shaded region indicates the time frame used to constrain the extent of mass stripping in the progenitor model. In the bottom panel, the model photospheric velocities are compared to the expansion velocity of SN~2021tsz, derived from the minima of the \ion{Fe}{ii} $\lambda$5169 absorption line.}
\label{2021tsz_wo_CSM}
\end{figure}

To determine the degree of mass stripping required to reproduce the observed decline rate of SN~2021tsz from 30 days post-explosion to t$_{\rm PT}$, we simulate explosions of progenitor models that have been stripped of 2\,M$_\odot$, 3\,M$_\odot$, and 4\,M$_\odot$ of their hydrogen envelope using \texttt{SNEC}. The $^{56}$Ni mass is fixed at 0.08 M$_\odot$, as estimated in Sect~\ref{sect:4.2}, and the outer extent of $^{56}$Ni mixing is set to a mass coordinate of 3\,M$_\odot$ as the different degrees of $^{56}$Ni mixing only has a subtle effect on the light curves (e.g. \citealt{Morozova2018}). The resulting bolometric light curves and photospheric velocities for an explosion energy of 1.3$\times$10$^{51}$ erg are shown in Fig.~\ref{2021tsz_wo_CSM}. This explosion energy is chosen because it optimally reproduces the observed photospheric velocities. The observed photospheric velocities correspond to \ion{Fe}{ii} $\lambda$5169 velocity. The blackbody-corrected luminosity of SN~2021tsz is estimated using \texttt{SuperBol} \citep{Nicholl2018}, based on $BgVri$ photometry, and is adopted as the observed bolometric luminosity. Since the blackbody approximation generally holds until the end of the recombination phase in Type II SNe, the luminosity estimates for the last two epochs of SN~2021tsz obtained with \texttt{SuperBol} may be somewhat over- or underestimated. However, as these two epochs are not included in our modelling, they do not affect our analysis. The observed luminosity during the photospheric phase lies between those of the 2\,M$_\odot$ and 3\,M$_\odot$ stripped models, while the 4\,M$_\odot$ stripped model declines too rapidly compared to the observed light curve. We therefore select the 2 and 3\,M$_\odot$ stripped progenitor models for further simulation of the light curve from explosion to t$_{\rm PT}$. The pre-SN configurations for these models are listed in Table~\ref{comp_model}. 

To model the light curve from the explosion to the end of the photospheric phase (<t$_{\rm PT}$), we attached CSM to the progenitor surface, assuming a spherical wind density profile: 
 ${\mathrm{\rho}}$ = Kr$^{-2}$, where r is the distance from the progenitor and K $\equiv$ $\dot{{\rm M}}$ /(4${\rm \pi}$v$_{{\rm wind}}$) is the mass-loading parameter (with $\dot{{\rm M}}$ being the mass-loss rate and v$_{{\rm wind}}$ the wind expansion velocity). Consequently, when CSM is integrated into the model, the mass loading parameter, K$_{\rm CSM}$, and the CSM extent, R$_{\rm CSM}$, become two additional free parameters. We generated a grid of models varying the explosion energy (E$_{\rm{exp}}$) from 1.0 to 2.0\,foe (1 foe = 10$^{51}$ erg) in increments of 0.1\,foe. The mass loading parameter (K$_{\rm CSM}$) was varied from 1 to 10\,$\times$\,10$^{17}$\,g\,cm$^{-1}$ in intervals of 1 $\times$ 10$^{17}$ g cm$^{-1}$, while the CSM radius (R$_{\rm CSM}$) spanned 2000 to 4000\,R$_\odot$ in 100 R$_\odot$ steps. To account for uncertainties in explosion time, the observed light curves were shifted by ±0.9\,d in steps of 0.3\,d, and this offset (t$_{\rm offset}$) was treated as a free parameter. The $^{56}$Ni mass was varied between 0.07 to 0.09 M$_\odot$ in steps of 0.01 M$_\odot$, with the $^{56}$ Ni mixing parameter fixed at 3 M$_\odot$. 

We employed the chi-square minimisation technique to obtain the model that simultaneously best fits the observed bolometric light curve from the explosion until t$_{\rm PT}$ and the observed photospheric velocity evolution.
The $\chi^2$ is computed as
\begin{equation}
\scalebox{1.04}{
$\chi^2 = \frac{1}{N}\sum\limits_{t \textless t_{\rm PT}} \left(\frac{X^\mathrm{obs}_\lambda(t) - X^\mathrm{model}_\lambda(t)}{\Delta X^\mathrm{obs}_\lambda(t)}\right)^2$,}
    \label{eq:chi2}
\end{equation}
where $X^\mathrm{obs}_\lambda$(t) are the observed luminosities and velocities at time t and $\Delta X^\mathrm{obs}_\lambda$(t) correspond to the associated errors,
$X^\mathrm{model}_\lambda$(t) corresponds to the model luminosities and velocities at time t, and $N$ is the number of observed data points. 

\begin{figure}
\begin{minipage}[h]{1.0\linewidth}
\centering
\includegraphics[scale=0.49, clip, trim={0cm 0.1cm 0 1.8cm}]{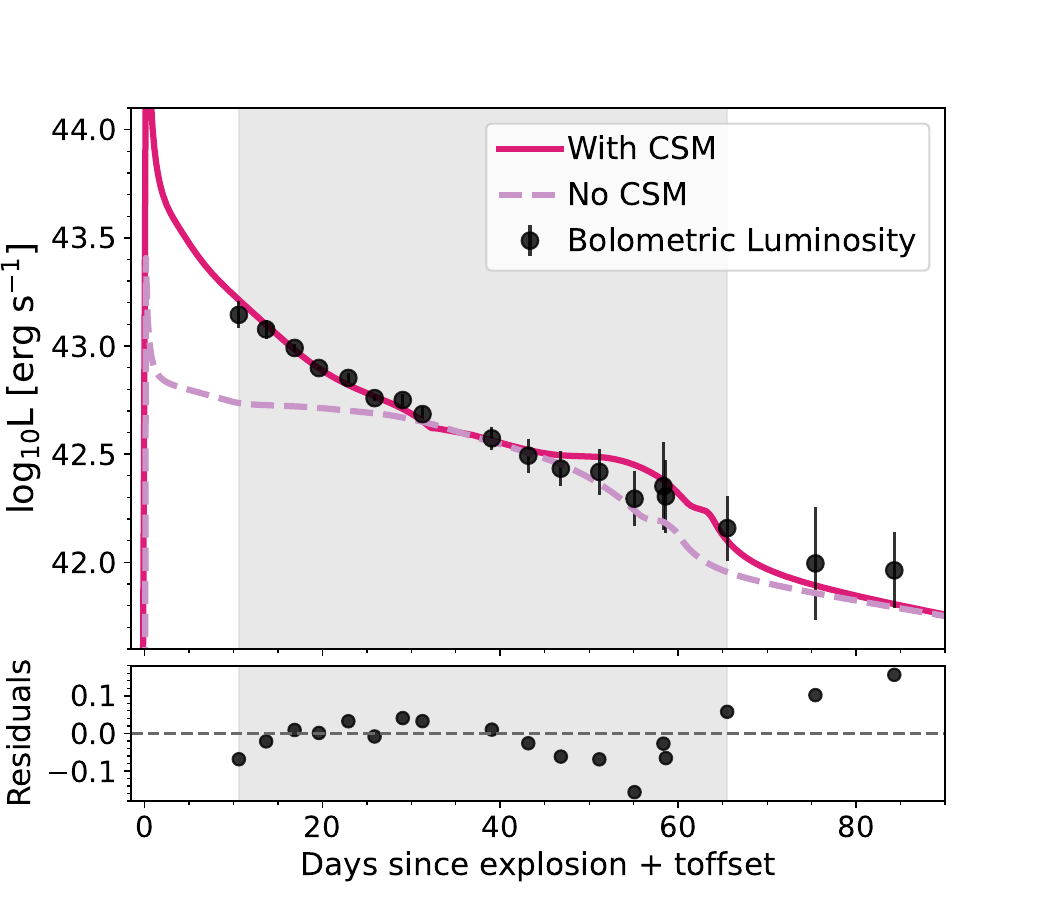}
\end{minipage}
\begin{minipage}[h]{1.0\linewidth}
\centering
\includegraphics[scale=0.40, clip, trim={0.4cm 0.2cm 0 0.2cm}]{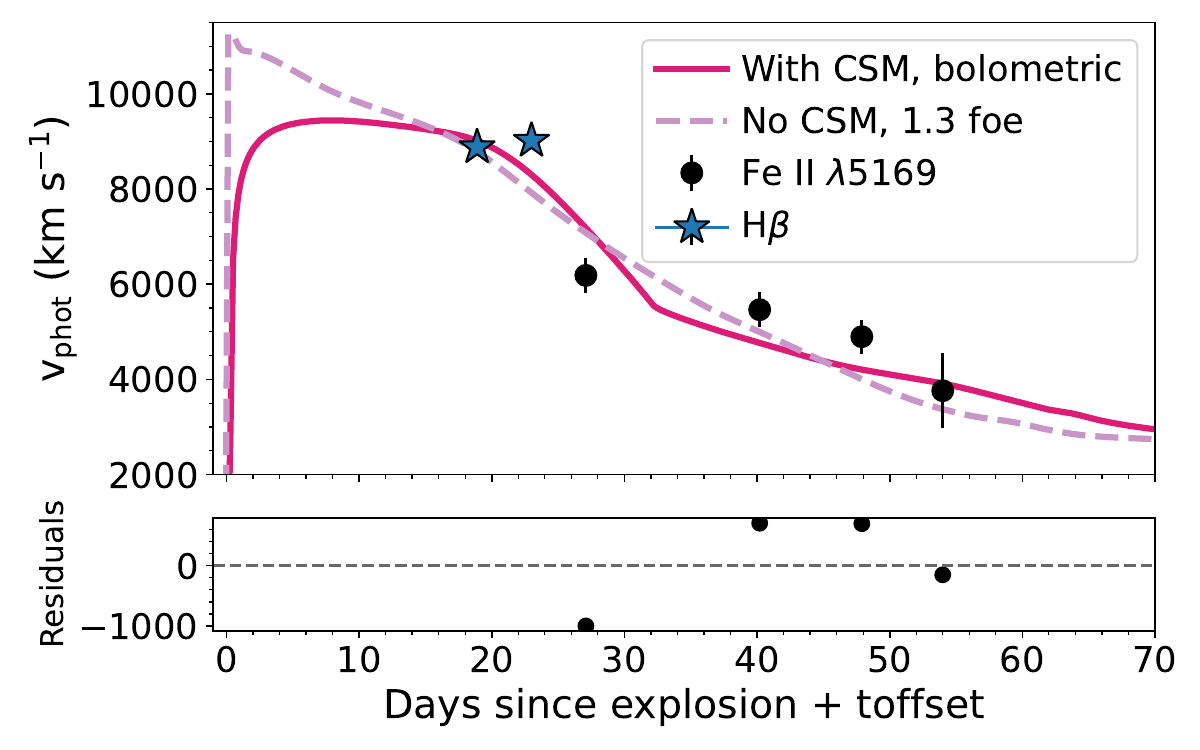}
\end{minipage}
\caption{Comparison of the observed and modelled bolometric light curve and photospheric velocity of SN~2021tsz. Top: Observed bolometric light curve alongside the best-fit SNEC model (solid line), obtained by simultaneously fitting the light curve and photospheric velocity up to 70 days. The dashed line shows the model without CSM. The shaded region indicates the fitting interval. Bottom: Model photospheric velocities plotted against observed values, derived from the minima of the \ion{Fe}{ii} $\lambda$5169 absorption line. An explosion epoch offset of -0.9 days is applied.}
\label{2021tsz_model_lum}
\end{figure}
 
\texttt{SNEC} applies bolometric corrections to derive synthetic multi-band light curves from the modelled bolometric luminosity. While these corrections are generally appropriate for normal Type II SNe, they may not reliably represent the photometric properties of fast-declining, short-photospheric phase events. Given this limitation, for SN~2021tsz we directly fit the observed bolometric light curve and expansion velocity, rather than relying on synthetic multi-band photometry, to determine the best-fit model parameters. The optimal model corresponds to a $3\,M_\odot$ stripped progenitor, characterised by a hydrogen envelope mass of $4\,M_\odot$, an explosion energy of 1.3\,foe, a CSM radial extent of $3100\,R_\odot$, and a mass-loading parameter of $7 \times 10^{17}$ g\,cm$^{-1}$. Fig.~\ref{2021tsz_model_lum} shows the evolution of the bolometric luminosity for the best-fit model compared to the observed light curve. For reference, we also plot a model without CSM, computed with the same progenitor and explosion parameters. The lower panel shows the evolution of the photospheric velocity for both models, overlaid with the observed expansion velocities derived from \ion{H}{$\beta$} and \ion{Fe}{ii} $\lambda$5169. At velocities above 8000 km s$^{-1}$, it is evident that \ion{H}{$\beta$} serves as a good proxy for the photospheric velocity when the photosphere resides in the outer hydrogen-rich layers — a finding consistent with previous studies of Type IIP SNe (e.g. \citealt{Takats2012}).

Based on the CSM parameters, the CSM mass can be written as
\begin{equation}
{\rm{M}_{\rm CSM}} = \int_{{\rm R}_{\rm{in}}}^{{\rm R}_{\rm CSM}} 4 \pi \rho_{\rm CSM}({\rm r}) {\rm r}^2 {\rm dr}
\end{equation}
where R$_{\rm{in}}$ is the inner CSM radius and R$_{\rm CSM}$ is the outer CSM radius.

Substituting $\rho_{\rm CSM}$(r) = ${\rm K}_{\rm CSM} {\rm r}^{-2}$ and replacing R$_{\rm{in}}$ with the progenitor radius as the CSM here is assumed to be attached to the progenitor, and integrating, we get 
$$ {\rm M}_{\rm CSM} = 4 \pi {\rm K}_{\rm CSM} ({\rm R}_{\rm CSM} - {\rm R}_{\star}). $$

Here, R$_{\star}$ represents the progenitor radius. From this expression, we calculate the total CSM mass to be 0.65\,M$_\odot$, using the best-fit parameter values: R$_{\star}$~=~991~R$_\odot$, R$_{\rm CSM}$ = 3100 R$_\odot$, and K$_{\rm CSM}$\,=\,7\,$\times$\,10$^{17}$\,g\,cm$^{-1}$. Such a high CSM mass was also obtained for SN~2013by (0.83\,M$_\odot$), a fast-declining Type II SN in \cite{Morozova2018}. Even for other fast declining Type II SNe, such as SNe~2013ej and 2014G, \cite{Hillier2019} predicted a CSM mass in the range of 0.5-1.0\,M$_\odot$, while for SN~1998S, \cite{Dessart2016} estimated a CSM mass of 0.4\,M$_\odot$. The best-fit parameters of the 2 and 3\,M$_\odot$ stripped progenitor models are summarised in Table~\ref{comp_model}. Finally, we note that the estimated values assume a symmetric ejecta–CSM configuration, although the possibility of asymmetry cannot be excluded.

While it is possible to estimate the mass-loss rate from the inferred CSM mass by assuming a reasonable wind velocity, we refrain from doing so in this work. This is because \texttt{SNEC} models primarily constrain the density structure of the CSM and are insensitive to the wind velocity in the pre-explosion configuration. Consequently, adopting a typical RSG wind speed of $\sim$10 km\,s$^{-1}$ would imply a relatively low mass-loss rate, whereas assuming higher wind velocities (e.g., $\sim$100 km\,s$^{-1}$) would yield significantly higher mass-loss rates, up to a few M$_\odot$\,yr$^{-1}$, in the period shortly preceding the explosion.

\section{Discussion}
\label{sec:7}
\subsection{Diversity among short-photospheric-phase Type II SNe}

\begin{figure}
\centering
\includegraphics[scale=0.41, clip, trim={1.7cm 0.4cm 0cm 0.2cm}]{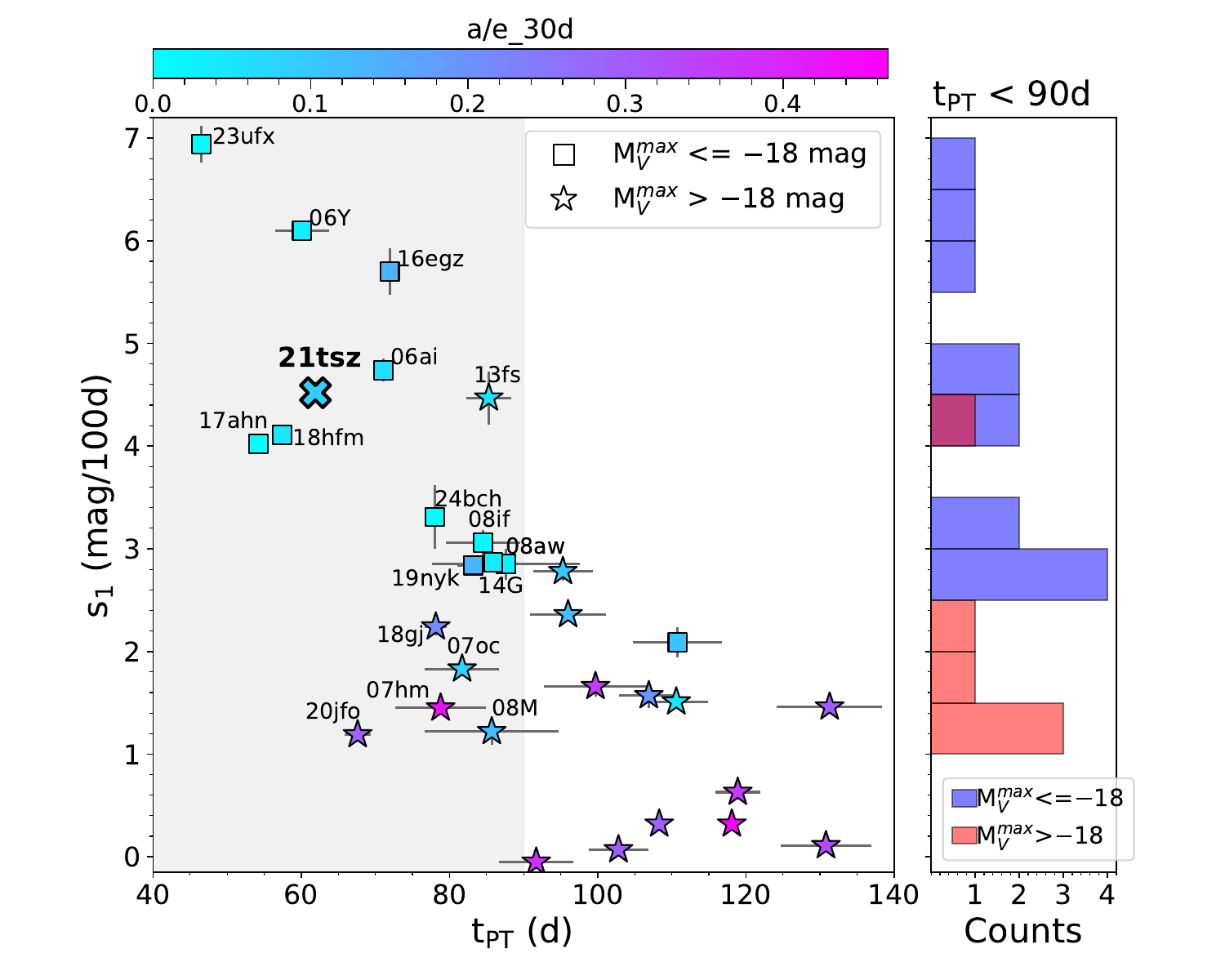}

\caption{Position of SN~2021tsz on the $V$-band slope (s$_1$) vs. t$_{\rm PT}$ plot, alongside other SNe II. SNe~II with photospheric phases shorter than 90 days are labelled, and this region is shaded in gray. SNe~II with peak magnitudes fainter than $-$18 mag are marked with stars ($\star$), while squares denote those brighter than $-$18 mag. The points are colour-coded by the absorption-to-emission ratio of \ion{H}{$\alpha$} at 30 days. Data are compiled from the literature \citep{Anderson2014, Valenti2016, Gutirrez2017, Dastidar2024}, along with independent estimates from this work. The right panel shows the histogram of the s$_1$ distribution for SNe~II with t$_{\rm PT} <$ 90 days, where those brighter than $-$18 mag are shown in blue and those fainter than $-$18 mag in red.}
\label{2021tsz_tpt_s2}
\end{figure}

To further examine the differences between fast- and slow-declining short-photospheric-phase Type II SNe, we plot $V$-band light curve parameters-- the initial decline rate from peak (s$_1$) versus the photospheric-phase duration (t$_{\rm PT}$)-- for our comparison sample, alongside SNe~II from \cite{Anderson2014}, \cite{Valenti2016}, \cite{Gutirrez2017}, \cite{Dastidar2024}, and this work in Fig.~\ref{2021tsz_tpt_s2}. The points are colour-coded by the absorption-to-emission ratio of \ion{H}{$\alpha$} at 30 days. Since the transition from the photospheric phase to the radioactive tail is not distinct in the $V$ band light curve of SN~2021tsz, the t$_{\rm PT}$ value estimated from the $r$ band is used here, while the decline rate is derived from the $V$ band. Type II SNe with t$_{\rm PT} < 90$ days are labelled in the figure, and this region is shaded in gray. Squares denote events with peak absolute magnitudes in $V$-band (M$_V^{max}$\footnote{In some events, the peak was not observed; in those cases, we use the brightest observed magnitude as M$_V^{\mathrm{max}}$.}) brighter than $-$18 mag, while stars ($\star$) represent events with peak absolute magnitudes fainter than $-$18 mag. The right panel shows the s$_1$ distribution for the subsample of SNe~II with t$_{\rm PT}$ $<$ 90 days. 

In general, slowly declining Type II SNe exhibit longer photospheric phases and higher \ion{H}{$\alpha$} absorption-to-emission ratios than their faster-declining counterparts, consistent with earlier population studies (e.g. \citealt{Gutirrez2017}). However, the early works of \citet{Anderson2014} and \citet{Valenti2016}, featured only rare instances of short-photospheric phase events, limited to SNe~1979C, 2006Y, and 2006ai. Based on the Type II SN sample from \citet{Gutirrez2018}, the fraction of short-photospheric phase events has been estimated to be approximately $\sim$4\% (3/78; \citealt{Hiramatsu2021}). Their rarity is also reflected in the models of \citet{Eldridge2018}, where only 4.7\% of simulated progenitors (30/637) produced short-photospheric phase Type II SN light curves. Over time, however, new discoveries have gradually built a small but growing sample, to which SN~2021tsz adds as a new member of this emerging population. 

\begin{figure}
\centering
\includegraphics[scale=0.55, clip, trim={5.6cm 2.2cm 0cm 3.6cm}]{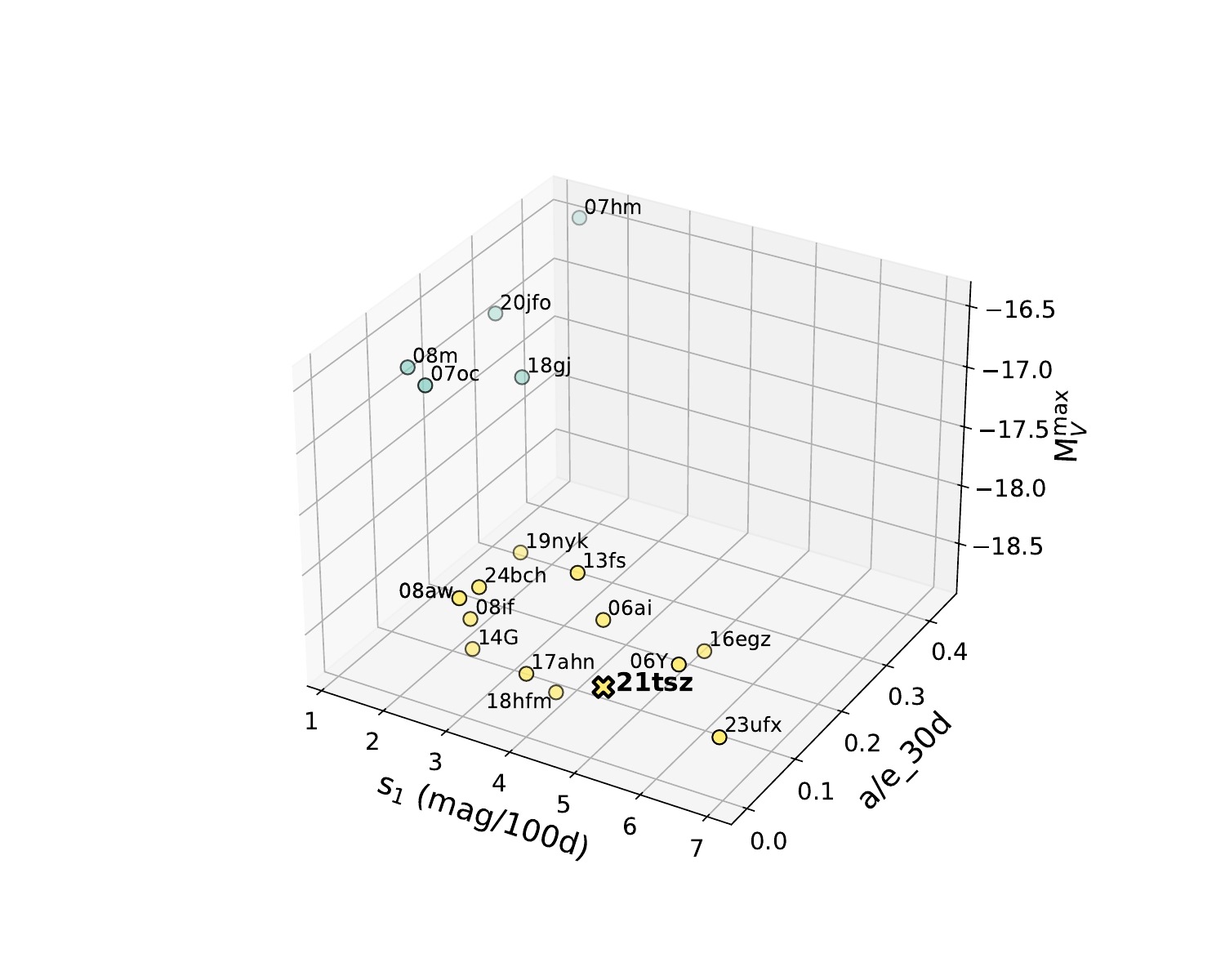}

\caption{Three-dimensional k-means clustering of the SN~II sample with plateau durations (t$_{\rm PT}$) $<$ 90 d, in the parameter space defined by the post-plateau decline rate (s$_1$), the spectral line strength ratio (a/e\_30d), and the absolute peak magnitude in $V$-band (M$_V^{\rm max}$). Different colours mark the two clusters identified by the algorithm, and the SN names are indicated for reference.}
\label{2021tsz_ae_s1_Mv_3D}
\end{figure}

In Fig.~\ref{2021tsz_tpt_s2}, we observe a bifurcation for $t_{\rm PT} <$ 90 days into two distinct categories: slow-declining, normal luminosity SNe~II (e.g., SNe~2020jfo and 2018gj), and fast-declining, luminous events such as SN~2021tsz, as also noted in the recent work of \citet{Ravi2025}. To explore this further, we applied a k-means clustering analysis to the subset of SNe~II with plateau durations ($t_{\rm PT}$) shorter than 90 days, using a three–dimensional parameter space defined by the post-peak decline rate (s$_1$), the \ion{H}{$\alpha$} absorption-to-emission ratio measured at 30 days (a/e\_30d), and the peak absolute magnitude in $V$ band (M$_V^{\rm max}$). Each object is assigned to one of the clusters derived from the algorithm, and the results are visualised in Fig.~\ref{2021tsz_ae_s1_Mv_3D} as a three–dimensional scatter plot with the cluster memberships distinguished by colour. 

The clustering analysis reveals two distinct groups. This bifurcation can be physically understood in terms of varying degrees of interaction with CSM. Stronger CSM interaction can prolong the apparent plateau duration, as densities remain high even after the recombination phase \citep{Dessart2016}. It can also increase the peak luminosity, since the ejecta–CSM shock efficiently converts kinetic energy into radiation \citep{Chevalier2017}. At the same time, CSM interaction tends to suppress the absorption in \ion{H}{$\alpha$}, because shock-generated photons originating in the outer, high-velocity ejecta can fill in the \ion{H}{$\alpha$} absorption trough \citep{Hillier2019, Dessart2022}. Thus, the observed separation of the sample into two clusters likely reflects the combined impact of CSM interaction on both the light-curve morphology and spectral line formation.

\subsection{Comparison with previous modelling works}
To contextualize our findings for SN~2021tsz, we compare them with previous modelling efforts on fast-declining, short-photospheric phase Type II SNe. In particular, we contrast our inferred progenitor and CSM properties with those from \citet{Hiramatsu2021} and related works, highlighting both consistencies and key differences. This helps to better understand the diversity of progenitor scenarios for such events.

SN~2021tsz exhibits a faster-declining light curve, with a luminous peak and a shorter photospheric phase compared to normal Type II SNe. Similar behaviour has been observed in Type II SNe~2006Y, 2006ai, 2016egz, and 2023ufx \citep{Hiramatsu2021, Ravi2025}, where progenitors with lower hydrogen envelope (H$_{\rm env}$) masses were invoked to explain the light curve properties through hydrodynamical modelling. In line with these findings, \citet{Hillier2019} demonstrated that a 15\,M$_\odot$ progenitor with a less massive hydrogen-rich envelope could naturally reproduce such fast-declining light curves, motivating our use of stripped progenitor models from \citet{Morozova2015} in this work.

Given the low metallicity of SN~2021tsz's host (derived from strong-line diagnostics in Sect.~\ref{sec:3.4}), strong mass loss via line-driven stellar winds would be disfavoured if RSG mass-loss rates are metallicity dependent \citep[e.g.,][]{Vink2000, Vink2001, Mokiem2007, Vink2021}. Instead, binary interactions likely played a key role in stripping the progenitor’s envelope, consistent with the high binary interaction fraction (up to 70\%) observed among massive stars in binary systems \citep{Sana2012}. Furthermore, population synthesis simulations by \citet{Zapartas2019} indicate that $\sim$30–50\% of all Type II SN progenitors are expected to have interacted with a companion prior to explosion, producing progenitors with a wide range of hydrogen-envelope masses even for similar ZAMS masses.

In terms of progenitor ZAMS mass estimates, \citet{Hiramatsu2021} proposed that ZAMS masses $>17$ M$_\odot$ could give rise to short-photospheric phase SNe, especially at sub-solar metallicities ($\sim$0.3 Z$_\odot$) where lower opacities yield more compact progenitors with radii around 480–580 R$_\odot$. On the other hand, \citet{Eldridge2018} suggested a broader ZAMS mass range of 7–25\,M$_\odot$ in binary evolution scenarios. In our modelling, we adopted a 15\,M$_\odot$ progenitor at solar metallicity, with envelope stripping occurring during the sub-giant branch phase. As pointed out by \citet{Dessart2019}, the light curve properties are primarily governed by the hydrogen-rich envelope and provide limited constraints on the helium core or ZAMS mass. Therefore, while our modelling suggests a stripped pre-SN star, it does not allow us to precisely constrain the progenitor's initial mass.

We find a pre-SN hydrogen envelope mass of $\sim$4\,M$_\odot$ corresponding to the 3\,M$_\odot$ H-stripped model, which is higher than the 0.91–2.12\,M$_\odot$ range estimated by \citet{Hiramatsu2021}. This discrepancy likely arises from structural differences in the progenitor models: \citet{Hiramatsu2021} assumed more compact progenitors, whereas we adopted an extended progenitor structure. Since the photospheric luminosity depends on the envelope’s density profile, compact progenitors require smaller envelope masses compared to extended ones to produce similar observed light curve features. Consequently, the pre-SN mass inferred by \citet{Hiramatsu2021} (7.1–8.5\,M$_\odot$) is also somewhat lower than that estimated for SN~2021tsz in this work (9.2\,M$_\odot$), although the helium core masses are comparable (5.1\,M$_\odot$ in this work and 5.4–6.9\,M$_\odot$ in \citealt{Hiramatsu2021}). Thus, adopting a solar-metallicity progenitor in this work, despite the host’s lower metallicity, may partly explain the higher pre-SN hydrogen envelope mass inferred here.

The CSM mass required to explain the early light curve excess in SN~2021tsz is higher than that inferred for the events studied by \citet{Hiramatsu2021} (0.1–0.3\,M$_\odot$), which had early excesses lasting only $\sim$10 days. In contrast, SN~2021tsz exhibited a much brighter and more prolonged early excess ($\sim$30 days), necessitating a denser and more massive CSM. Assuming a standard RSG wind velocity (10 km s$^{-1}$), this material would have been ejected $\sim$4.6 years before core collapse, while a faster ‘super-wind’ (100 km s$^{-1}$; e.g. \citet{Smith2023}) would place the onset within $\sim$6 months of explosion.

Our modelling thus indicates the presence of a dense CSM close to the progenitor, likely formed during the final evolutionary stages. While binary interactions remain the most plausible channel for producing such a dense environment, \cite{Zapartas2019} argued that binary interactions occurring shortly before core collapse are statistically rare unless triggered by late-stage stellar instabilities, such as envelope inflation induced by nuclear burning instabilities \citep[e.g.,][]{Mcley2014, Smith2014} or wave-driven mass loss \citep[e.g.,][]{Fuller2017}. Such instabilities could also independently drive CSM formation in single-star progenitors. Alternatively, colliding winds in close binaries could also generate dense boundary layers that mimic mass loss from single stars \citep{Kochanek2019}. The precise origin of the CSM in SN~2021tsz, therefore, remains uncertain, but it most likely reflects a complex interplay between binary evolution and late-stage stellar instabilities. Finally, we note the absence of narrow emission lines in our spectra, despite evidence for a dense, nearby CSM. Such features may have disappeared within the first week, when no spectroscopy was obtained. Alternatively, in high-density environments (as inferred here), the mean free path of recombination photons that would normally produce IIn-like narrow lines can be much smaller than the CSM’s radial extent \citep{Dessart2017}, thereby suppressing their appearance in this event.

\section{Summary}\label{sec:8}

With the advent of untargeted surveys, an increasing number of SNe are being discovered in low-luminosity, dwarf galaxies (e.g. \citealt{Gutirrez2018}). We present the analysis of one such event, SN~2021tsz, a luminous, fast-declining Type II SN. It reaches a peak absolute magnitude of $-$18.88 mag in the $r$-band and exhibits a rapid decline of 4.0 mag (100d)$^{-1}$. The photospheric phase is notably short, lasting about 63 days, followed by a 0.97 mag drop to the radioactive decay tail in the $r$-band. The host galaxy, determined through SED analysis and strong-line diagnostics, is a low-metallicity dwarf galaxy (M$_\star$ = 1.2$\times$10$^8$ M$_\odot$, Z = 0.3~Z$_\odot$). 

Spectroscopically, SN~2021tsz shows a \ion{H}{$\alpha$} profile with a weak absorption component, indicative of strong CSM interaction \citep{Hillier2019}, and consistent with brighter, faster-declining Type II SNe \citep{Gutirrez2014}. \ion{Fe}{ii} features appear from 28 day onwards, while earlier spectra are dominated by a blue continuum, likely due to interaction. The spectroscopic properties of SN~2021tsz resemble those of other fast-declining short photospheric phase Type II SNe, whereas slow-declining events display stronger hydrogen and metal absorption lines. Our clustering analysis shows that these groups are separated in a three dimensional parameter space defined by decline rate from peak, \ion{H}{$\alpha$} absorption-to-emission ratio, and $V$-band peak absolute magnitude, which we interpret as arising from different degrees of CSM interaction.

Hydrodynamical modelling of the light curve indicates that CSM interaction dominates the luminosity during the first 30 days. To reproduce the short photospheric phase, partially stripped pre-SN models are used. Considering the host's low metallicity, strong winds would be unlikely, and binary interactions are a more probable cause of the stripping. The modelling also points to late-stage mass loss forming a dense shell near the progenitor, potentially arising from nuclear burning instabilities or radiatively cooled boundary layers formed by colliding winds in a binary system. The best-fit model suggests a pre-SN configuration with a hydrogen envelope mass of 4~M$_\odot$, surrounded by compact, dense CSM extending up to $\sim$3100 R$_\odot$, with a total CSM mass of 0.6 M$_\odot$. The explosion energy required to reproduce the photospheric phase luminosity and velocity is 1.3 $\times$ 10$^{51}$ erg.

The low-metallicity environment and the derived high CSM mass from hydrodynamical modelling strongly support a binary progenitor system for SN~2021tsz. The degree of envelope stripping implied by its short photospheric phase and light curve evolution is difficult to achieve through single-star evolution, especially at low metallicity where line-driven winds are weak. Binary interaction thus emerges as the most likely mechanism responsible for the reduced hydrogen envelope mass. Furthermore, the scarcity of such events suggests that the conditions required for their formation are rare. If binary evolution is indeed the dominant channel, it more often leads to progenitors with even lower hydrogen envelope masses, favouring the production of SNe IIb instead. This could naturally explain the rarity of short-photospheric phase Type II SN. In this context, SN~2021tsz represents a transitional case, bridging the gap between classical Type IIP and stripped-envelope Type IIb explosions.

\section*{Data availability}
All spectra are publicly available\footnote{\href{https://www.wiserep.org/object/18833}{https://www.wiserep.org/object/18833}} on the WISeREP interface \citep{Yaron2012}.

\begin{acknowledgements}
      We express our gratitude to the anonymous referee for providing us with valuable suggestions and scientific insights that enhanced the quality of the paper. This work makes use of data from the Las Cumbres Observatory global network of telescopes. The LCO group is supported by NSF grants AST-1911151 and AST-1911225. G.P. acknowledges support from the National Agency for Research and Development (ANID) through the Millennium Science Initiative Program – ICN12\_009. M.S. acknowledges financial support provided under the National Post Doctoral Fellowship (N-PDF; File Number: PDF/2023/002244) by the Science \& Engineering Research Board (SERB), Anusandhan National Research Foundation (ANRF), Government of India. C.P.G. acknowledges financial support from the Secretary of Universities and Research (Government of Catalonia) and by the Horizon 2020 Research and Innovation Programme of the European Union under the Marie Sk\l{}odowska-Curie and the Beatriu de Pin\'{o}s 2021 BP 00168 programme, from the Spanish Ministerio de Ciencia e Innovaci\'{o}n (MCIN) and the Agencia Estatal de Investigaci\'{o}n (AEI) 10.13039/501100011033 under the PID2023-151307NB-I00 SNNEXT project, from Centro Superior de Investigaciones Cient\'{i}ficas (CSIC) under the PIE project 20215AT016 and the program Unidad de Excelencia Mar\'{i}a de Maeztu CEX2020-001058-M, and from the Departament de Recerca i Universitats de la Generalitat de Catalunya through the 2021-SGR-01270 grant. AK is supported by the UK Science and Technology Facilities Council (STFC) Consolidated grant ST/V000853/1. We acknowledge Wiezmann Interactive Supernova data REPository \url{http://wiserep.weizmann.ac.il} (WISeREP, \citealt{Yaron2012}). This research has made use of the NASA/IPAC Extragalactic Database (NED), which is operated by the Jet Propulsion Laboratory, California Institute of Technology, under contract with the National Aeronautics and Space Administration. 
\end{acknowledgements}

\bibliographystyle{aa} 
\bibliography{aa56058-25}

\clearpage
\begin{appendix}
\section{Additional tables and figures}
\begin{figure}[H]
\centering
\includegraphics[scale=0.35, clip, trim={0.25cm 1.5cm 0 3.50cm}]{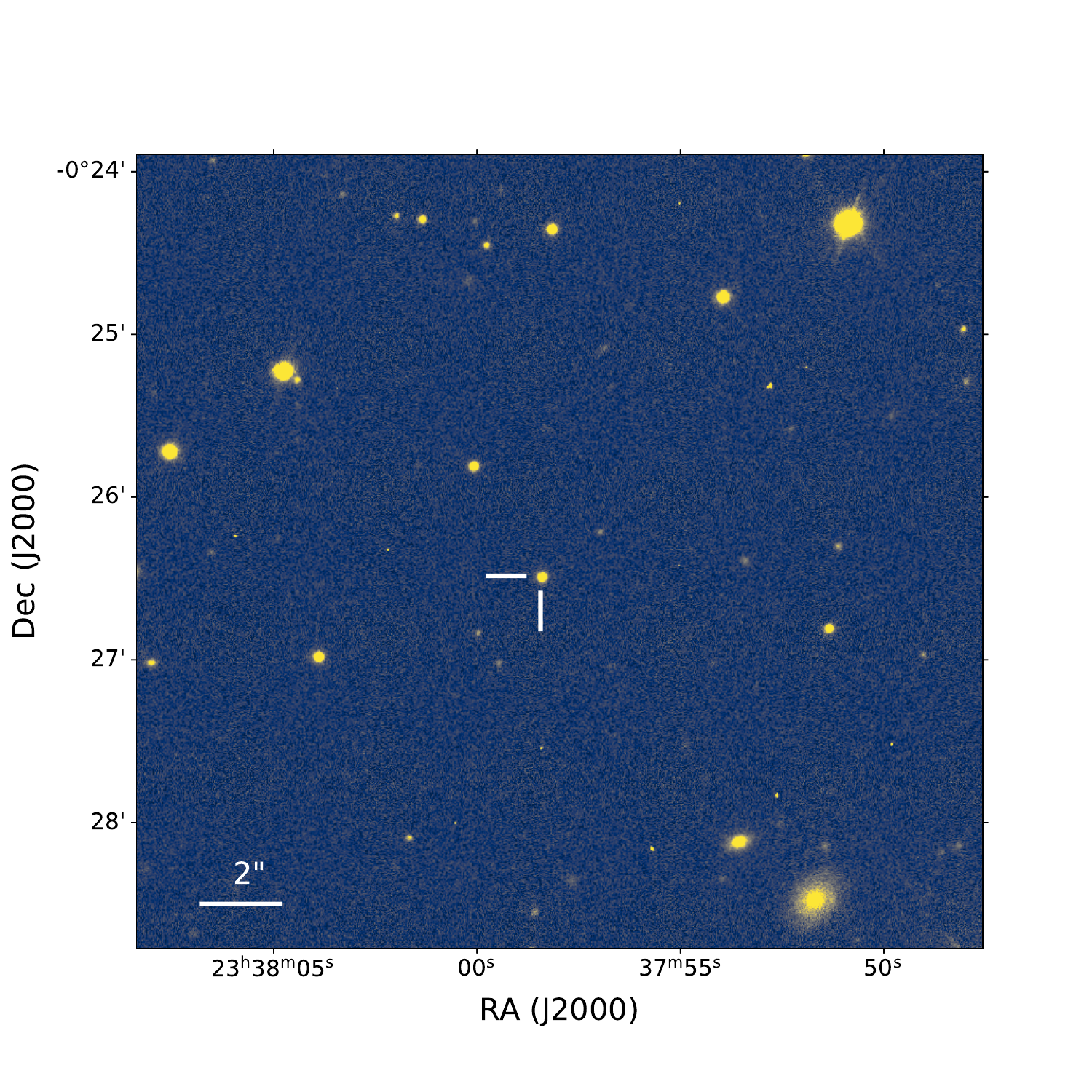}
\caption{A 200 s $r$-band image of the SN field obtained with the 1 m LCO telescope on 2021 July 30. The location of SN~2021tsz in the low-luminosity host galaxy SDSS J233758.39–002629.6 is marked. The SN is located very close to the galaxy nucleus—0.12$^{\prime\prime}$ S and 0.00$^{\prime\prime}$ W of the centre. As the host is a low-luminosity dwarf galaxy, the SN outshines it during the early phases, and only the SN is visible in this image.}
\label{2021tsz_fits}
\end{figure}
\begin{table}[H]
    \centering
    \footnotesize
    \renewcommand{\arraystretch}{1.2}
\setlength{\tabcolsep}{4.5pt}
    \caption{Basic information of SN~2021tsz and its host galaxy.}
    \scalebox{0.84}{
    \begin{tabular}{ll} 
		\hline\hline
            R.A.$^1$ & 23$^{\rm h}$37$^{\rm m}$58$^{\rm s}$.39\\
            Dec.$^1$ & $-$00$^\circ$26$^\prime$29$^{\prime\prime}$.51\\
            Discovery$^1$ & JD 2459414.9\\
            Explosion epoch$^{2}$  & JD 2459414.0 ± 0.9\\
            Redshift$^2$ &  0.0366 $\pm$ 0.0020\\
            E(B$-$V)$_{\rm MW}^{1}$ & 0.0297±0.0018 mag\\
            E(B$-$V)$_{\rm host}^2$ & 0 mag\\
            Distance$^{2}$ & 166.7 ± 9.8 Mpc \\
            Distance modulus$^{2}$ & 36.11 ± 0.13 mag\\
            \vspace*{0.1cm}
            Peak M$_g^2$       & $-$18.96 ± 0.13 mag\\
            \hline\hline
            \vspace*{0.1cm}
            Host Galaxy & SDSS J233758.38-002629.6 \\
            \hline
            GALEX FUV$^3$  & 23.31 ± 0.34 mag\\
            GALEX NUV$^3$  & 22.99 ± 0.20 mag\\
            SDSS u$^4$          & 21.47 ± 0.17 mag\\
            SDSS g$^4$          & 20.47 ± 0.03 mag\\
            SDSS r$^4$          & 20.11 ± 0.03 mag\\
            SDSS i$^4$          & 19.91 ± 0.03 mag\\
            SDSS z$^4$          & 19.60 ± 0.09 mag\\
            12 + log(O/H)$^{2}$ & 8.17 ± 0.03 dex ($\sim$0.3 Z$_\odot$)\\
            SFR$^{2,5}$ (using \ion{H}{$\alpha$} flux) & 0.0090 ± 0.0003 M$_\odot$ yr$^{-1}$\\
            SFR$^{2,6}$ (using FUV mag) & 0.0096 ± 0.0022 M$_\odot$ yr$^{-1}$\\
            M$_\star$$^{2,7}$  & (2.1 $\pm$ 0.3) $\times$ 10$^{8}$ M$_\odot$\\
            M$_\star$$^{2,8}$  & (1.2 $\pm$ 0.6) $\times$ 10$^{8}$ M$_\odot$\\
            SFR$^{2,8}$        & 0.13 $\pm$ 0.08 M$_\odot$ yr$^{-1}$\\
            log(sSFR)$^{2,8}$ (SFR/M$_\star$)      & $\sim$ $-$8.9 $\pm$ 0.4 yr$^{-1}$\\
		\hline
    \end{tabular}}
    \newline
	\noindent
\tablefoot{$^{1}${TNS}, $^{2}${This work}, $^{3}${https://galex.stsci.edu/GR6/}, $^4${https://skyserver.sdss.org/dr18/}, $^{5}${Eqn 2, \citealt{Kennicutt1998}}, $^{6}${Eqn 3, \citealt{Karachentsev2013}}, $^7${\cite{Bell2003}}, $^8${CIGALE, \citealt{Boquien2019}}}
	\label{Tab5:SN2021tsz}
\end{table}
\begin{table}
\centering
\caption{Log of spectroscopic observations.\label{Tab4:sn_spectra_log}}
\setlength{\tabcolsep}{3pt}
\renewcommand{\arraystretch}{1.1}
\scalebox{0.8}{
\begin{tabular}{c c c c c c}
\hline 
UT Date     & JD & Phase\tablefootmark{$\dagger$} & Instrument/ & Exposure & SNR \\
(yyyy-mm-dd) &    & (d) &  Telescope        & time (s) & (around \ion{H}{$\alpha$})\\
\hline
\noalign{\smallskip}
2021-07-26.4 & 2459422.0 & 8.0 	 & FLOYDS/FTN & 3600 & 16.6\\
2021-07-30.6 & 2459426.1 & 12.1  & FLOYDS/FTS & 3600 & 16.5\\
2021-08-02.9 & 2459428.8 & 14.9  & EFOSC/NTT  & 300  & 19.7\\
2021-08-06.4 & 2459432.9 & 18.9  & FLOYDS/FTN & 3600 & 24.6\\
2021-08-10.5 & 2459437.0 & 23.0  & FLOYDS/FTN & 3600 & 29.5\\
2021-08-15.6 & 2459442.1 & 28.1  & FLOYDS/FTN & 3600 & 20.3\\
2021-08-22.5 & 2459449.0 & 35.0  & FLOYDS/FTN & 3600 & 9.4\\
2021-08-28.7 & 2459455.2 & 41.2  & FLOYDS/FTS & 3600 & 11.7\\
2021-09-05.4 & 2459462.9 & 48.9  & FLOYDS/FTN & 3600 & 26.8\\
2021-09-11.5 & 2459469.0 & 55.0  & FLOYDS/FTS & 3600 & 4.6\\
2021-09-17.4 & 2459475.0 & 61.0  & FLOYDS/FTN & 3600 & 9.3\\
\hline
\end{tabular}
}
\newline
\tablefoot{$\dagger${\scriptsize{since the explosion epoch t$_0$\,=\,JD\,2459414.0}}}
\end{table}
\begin{table}
 \caption{ZTF $g$, $r$ and $i$ band forced photometry of SN~2021tsz.}
 \label{Tab2_ZTFphot}
 \setlength{\tabcolsep}{3pt}
 \scalebox{0.74}{
 \begin{tabular}{cccccc}
 \hline
UT Date & JD  & Phase\tablefootmark{$\dagger$} & $g$         & $r$  & $i$\\
(yyyy-mm-dd) &     & (days)          & (mag)  & (mag) & (mag)\\
\hline
2021-07-19.4 & 2459414.9 & 0.9 & 18.263 ± 0.070 & -- & -- \\
2021-07-21.4 & 2459416.9 & 2.9 & -- & 17.774 ± 0.047 & -- \\
2021-07-22.4 & 2459417.9 & 3.9 & -- & -- & 17.766 ± 0.042 \\
2021-07-23.4 & 2459418.9 & 4.9 & 17.457 ± 0.053 & 17.600 ± 0.062 & -- \\
2021-07-30.3 & 2459425.9 & 11.9 & 17.454 ± 0.046 & -- & 17.468 ± 0.036 \\
2021-08-02.5 & 2459429.0 & 15.1 & -- & 17.405 ± 0.043 & -- \\
2021-08-03.3 & 2459429.8 & 15.8 & -- & -- & 17.460 ± 0.027 \\
2021-08-08.4 & 2459434.9 & 20.9 & 17.886 ± 0.051 & 17.611 ± 0.052 & 17.512 ± 0.026 \\
2021-08-10.4 & 2459436.9 & 22.9 & 17.982 ± 0.047 & -- & -- \\
2021-08-13.4 & 2459439.9 & 25.9 & -- & 17.846 ± 0.06 & 17.754 ± 0.035 \\
2021-08-16.3 & 2459442.8 & 28.8 & 18.372 ± 0.089 & -- & -- \\
2021-08-18.4 & 2459444.9 & 30.9 & 18.518 ± 0.075 & -- & -- \\
2021-08-21.3 & 2459447.8 & 33.8 & -- & -- & 18.067 ± 0.138 \\
2021-08-22.3 & 2459448.8 & 34.8 & 18.730 ± 0.198 & 18.236 ± 0.105 & -- \\
2021-08-26.3 & 2459452.8 & 38.8 & 18.936 ± 0.098 & -- & -- \\
2021-08-28.3 & 2459454.9 & 40.9 & 18.998 ± 0.105 & 18.388 ± 0.064 & -- \\
2021-08-29.3 & 2459455.8 & 41.8 & -- & -- & 18.263 ± 0.045 \\
2021-09-01.3 & 2459458.8 & 44.8 & 19.168 ± 0.098 & 18.479 ± 0.061 & 18.364 ± 0.054 \\
2021-09-03.4 & 2459460.8 & 46.8 & -- & 18.493 ± 0.060 & -- \\
2021-09-04.3 & 2459461.8 & 47.8 & -- & -- & 18.455 ± 0.047 \\
2021-09-05.3 & 2459462.8 & 48.9 & 19.349 ± 0.112 & 18.560 ± 0.058 & -- \\
2021-09-07.3 & 2459464.8 & 50.9 & 19.511 ± 0.117 & 18.649 ± 0.073 & 18.633 ± 0.048 \\
2021-09-09.4 & 2459466.9 & 52.9 & -- & 18.651 ± 0.068 & -- \\
2021-09-10.3 & 2459467.8 & 53.8 & -- & -- & 18.612±0.048 \\
2021-09-11.3 & 2459468.8 & 54.9 & 19.694 ± 0.116 & 18.792 ± 0.068 & -- \\
2021-09-13.3 & 2459470.8 & 56.9 & 19.849 ± 0.136 & 18.806 ± 0.069 & 18.781 ± 0.058 \\
2021-09-15.4 & 2459472.9 & 58.9 & 20.032 ± 0.215 & 18.917 ± 0.092 & -- \\
2021-09-16.4 & 2459473.9 & 59.9 & -- & -- & 18.846 ± 0.067 \\
2021-09-17.3 & 2459474.8 & 60.8 & 20.104 ± 0.313 & 18.977 ± 0.098 & -- \\
2021-09-22.3 & 2459479.8 & 65.8 & -- & 19.320 ± 0.198 & 19.101 ± 0.192 \\
2021-09-27.3 & 2459484.8 & 70.8 & -- & 19.627 ± 0.134 & 19.793 ± 0.133 \\
2021-09-30.3 & 2459487.8 & 73.8 & -- & -- & 19.706 ± 0.159 \\
2021-10-01.3 & 2459488.8 & 74.8 & 20.822 ± 0.309 & 19.725 ± 0.140 & -- \\
2021-10-07.3 & 2459494.8 & 80.8 & -- & -- & 19.897 ± 0.287 \\
2021-10-10.3 & 2459497.8 & 83.8 & -- & 19.810 ± 0.156 & -- \\
2021-10-15.3 & 2459502.8 & 88.8 & -- & 19.948 ± 0.202 & 19.994 ± 0.256 \\
2021-10-22.3 & 2459509.8 & 95.8 & -- & 19.846 ± 0.186 & 19.790 ± 0.185 \\
2021-10-25.2 & 2459512.7 & 98.7 & -- & 19.860 ± 0.135 & -- \\
2021-10-30.3 & 2459517.8 & 103.8 & -- & -- & 20.111 ± 0.183 \\
2021-11-02.2 & 2459520.7 & 106.8 & -- & 19.906 ± 0.226 & 20.138 ± 0.207 \\
2021-11-04.2 & 2459522.7 & 108.7 & -- & 20.090 ± 0.173 & -- \\
2021-11-05.2 & 2459523.7 & 109.7 & -- & -- & 20.396 ± 0.215 \\
2021-11-06.2 & 2459524.7 & 110.7 & -- & 19.945 ± 0.145 & -- \\
2021-11-08.2 & 2459526.7 & 112.7 & -- & 20.228 ± 0.182 & 20.021 ± 0.171 \\
\hline
 \end{tabular}}
 \tablefoot{$\dagger${\scriptsize since the explosion epoch t$_0$\,=\,JD\,2459414.0}}
\end{table}
\begin{table*}
\caption{The magnitudes of SN~2021tsz in $BVgri$ filters. The $BV$-band photometry are given in Vega magnitudes and $gri$-band data are presented in AB magnitudes.}\label{Tab1:2021tsz_phot}
\renewcommand{\arraystretch}{1.1}
\setlength{\tabcolsep}{4.5pt}
\centering
\scalebox{0.87}{
\begin{tabular}{ccclllll}
\hline
 Date  &         JD & Phase \tablefootmark{$\dagger$} &    $B$ (mag) &   $V$ (mag) &     $g$ (mag) &       $r$ (mag)  &        $i$ (mag) \\
  (yyyy-mm-dd) &         & (d) &  &     &    &          &         \\
\hline
2021-07-30.1 & 2459425.6 & 11.6 & 17.557 ± 0.020 & 17.263 ± 0.023 & 17.265 ± 0.017 & 17.316 ± 0.022 & 17.329 ± 0.024 \\
2021-08-02.2 & 2459428.7 & 14.7 & 17.734 ± 0.017 & 17.348 ± 0.017 & 17.409 ± 0.007 & 17.343 ± 0.014 & 17.341 ± 0.018 \\
2021-08-05.4 & 2459431.9 & 17.9 & 17.889 ± 0.016 & 17.527 ± 0.015 & 17.629 ± 0.006 & 17.525 ± 0.009 & 17.537 ± 0.009 \\
2021-08-08.1 & 2459434.6 & 20.6 & 18.182 ± 0.019 & 17.653 ± 0.017 & 17.880 ± 0.008 & 17.602 ± 0.009 & 17.558 ± 0.012 \\
2021-08-11.4 & 2459437.9 & 23.9 & 18.313 ± 0.018 & 17.815 ± 0.016 & 17.968 ± 0.006 & 17.738 ± 0.009 & 17.690 ± 0.020 \\
2021-08-14.4 & 2459440.9 & 26.9 & 18.585 ± 0.019 & 18.014 ± 0.018 & 18.235 ± 0.007 & 17.923 ± 0.011 & 17.874 ± 0.01 \\
2021-08-17.5 & 2459444.0 & 30.0 & 18.543 ± 0.047 & 18.082 ± 0.031 & 18.235 ± 0.019 & 17.971 ± 0.015 & 18.003 ± 0.032 \\
2021-08-19.8 & 2459446.2 & 32.2 & 18.856 ± 0.030 & 18.186 ± 0.031 & 18.433 ± 0.021 & 18.070 ± 0.019 & 18.029 ± 0.024 \\
2021-08-27.5 & 2459454.0 & 40.0 & 19.277 ± 0.036 & 18.515 ± 0.037 & 18.830 ± 0.030 & 18.261 ± 0.027 & 18.210 ± 0.047 \\
2021-08-31.6 & 2459458.1 & 44.1 & 19.630 ± 0.026 & 18.734 ± 0.028 & 19.141 ± 0.015 & 18.445 ± 0.016 & 18.448 ± 0.026 \\
2021-09-04.3 & 2459461.8 & 47.8 & 19.923 ± 0.033 & 18.884 ± 0.030 & 19.404 ± 0.021 & 18.557 ± 0.038 & 18.681 ± 0.075 \\
2021-09-08.6 & 2459466.1 & 52.1 & 20.172 ± 0.029 & 19.071 ± 0.027 & 19.587 ± 0.020 & 18.672 ± 0.021 & 18.732 ± 0.032 \\
2021-09-15.8 & 2459473.4 & 59.3 & -- & -- & -- & 18.862 ± 0.051 & -- \\
2021-10-02.9 & 2459490.4 & 76.4 & 21.351 ± 0.094 & 20.403 ± 0.074 & 21.142 ± 0.070 & 19.664 ± 0.040 & 19.876 ± 0.063 \\
2021-10-10.1 & 2459497.6 & 83.6 & -- & -- & 21.158 ± 0.218 & -- & -- \\
2021-10-11.8 & 2459499.3 & 85.3 & 21.614 ± 0.352 & 20.526 ± 0.124 & 21.148 ± 0.112 & 19.816 ± 0.084 & 19.92 ± 0.138 \\
2021-10-27.5 & 2459515.0 & 101.0 & -- & -- & -- & 19.864 ± 0.046 & 20.273 ± 0.203 \\
2021-11-06.5 & 2459525.0 & 111.0 & 21.993±0.186 & 20.895 ± 0.131 & 21.973 ± 0.081 & 20.030 ± 0.044 & 20.389 ± 0.091 \\
2021-11-30.2 & 2459548.6 & 134.6 & 22.653 ± 0.266 & 21.455 ± 0.144 & 21.834 ± 0.077 & 20.230 ± 0.047 & 20.681 ± 0.178 \\
2021-12-26.1 & 2459574.6 & 160.6 & 22.641 ± 0.507 & 21.497 ± 0.164 & 22.362 ± 0.192 & 20.457 ± 0.108 & 21.255 ± 0.222 \\
\hline
\end{tabular}}
\newline
\tablefoot{$\dagger$since the explosion epoch t$_0$\,=\,JD\,2459414.0}
\end{table*}
\begin{table}
 \caption{ATLAS $o$ and $c$ band forced photometry of SN~2021tsz.}
 \label{Tab3:atlas_phot}
 \setlength{\tabcolsep}{3.5pt}
 \scalebox{0.9}{
 \begin{tabular}{ccccc}
 \hline
UT Date & JD  & Phase\tablefootmark{$\dagger$} & $o$         & $c$\\
(yyyy-mm-dd) &     & (days)          & (mag)  & (mag)\\
\hline
2021-07-19.5 & 2459415.0 &  1.0 & -- & 18.402 ± 0.038 \\
2021-07-25.5 & 2459421.0 &  7.0 & 17.548 ± 0.039 & -- \\
2021-07-29.6 & 2459425.1 & 11.1 & 17.296 ± 0.083 & --\\
2021-07-31.6 & 2459427.1 & 13.1 & 17.443 ± 0.036 & --\\
2021-08-02.5 & 2459429.0 & 15.0 & 17.513 ± 0.015 & --\\
2021-08-06.5 & 2459433.0 & 19.0 & -- & 17.713 ± 0.017 \\
2021-08-08.5 & 2459435.0 & 21.0 & 17.586 ± 0.022 & --\\
2021-08-10.5 & 2459437.0 & 23.0 & -- & 17.924 ± 0.018 \\
2021-08-12.5 & 2459439.0 & 25.1 & 17.768 ± 0.021 & --\\
2021-08-14.5 & 2459441.0 & 27.0 & -- & 18.050 ± 0.023 \\
2021-08-16.5 & 2459443.0 & 29.0 & 17.958 ± 0.027 & --\\
2021-08-18.5 & 2459445.0 & 31.0 & 18.104 ± 0.037 & --\\
2021-08-21.4 & 2459447.9 & 33.9 & 18.236 ± 0.066 & --\\
2021-08-26.5 & 2459453.0 & 39.0 & 18.364 ± 0.048 & --\\
2021-08-28.5 & 2459455.0 & 41.0 & 18.472 ± 0.056 & --\\
2021-08-30.5 & 2459457.0 & 43.0 & 18.426 ± 0.037 & --\\
2021-09-01.5 & 2459459.0 & 45.0 & 18.454 ± 0.052 & --\\
2021-09-03.5 & 2459461.0 & 47.0 & 18.605 ± 0.044 & --\\
2021-09-05.5 & 2459463.0 & 49.0 & -- & 19.109 ± 0.063 \\
2021-09-07.5 & 2459465.0 & 51.0 & 18.636 ± 0.045 & --\\
2021-09-09.5 & 2459467.0 & 53.0 & -- & 19.344 ± 0.078 \\
2021-09-13.4 & 2459470.9 & 56.9 & 18.789 ± 0.056 & 19.419 ± 0.175 \\
2021-09-17.4 & 2459474.9 & 60.9 & 18.993 ± 0.115 & --\\
2021-09-27.4 & 2459484.9 & 70.9 & 19.544 ± 0.172 & --\\
2021-09-29.4 & 2459486.9 & 72.9 & 19.658 ± 0.121 & --\\
2021-10-01.4 & 2459488.9 & 74.9 & 19.804 ± 0.175 & -- \\
2021-10-03.4 & 2459490.9 & 76.9 & 19.720 ± 0.133 & -- \\
2021-10-07.4 & 2459494.9 & 80.9 & 19.868 ± 0.170 & -- \\
\hline
 \end{tabular}}
\tablefoot{$\dagger${\scriptsize since the explosion epoch t$_0$\,=\,JD\,2459414.0}}
\end{table}
\begin{table}
\caption{Light curve slopes in different phases.}
\label{Tab:LC_slopes}
\renewcommand{\arraystretch}{1.1}
\setlength{\tabcolsep}{4.5pt}
\footnotesize
\centering
\begin{tabular}{cccccc}
\hline
Band & t$_{\rm start}$ & t$_{\rm stop}$ & slope      & M$_{\rm peak}$ & t$_{\rm peak}$\\
     &     (d)     &  (d)       & (mag (100d)$^{-1}$) &   (mag)    & (d) \\
\hline
$B$    & 12.2 &   52.7 &	6.44 $\pm$ 0.06      \\
       & 52.7 &   77.0 &    4.94 $\pm$ 0.36 \\
       & 85.9 &  161.2 &    1.50 $\pm$ 0.75 \\
\hline
$g$    & 12.2 &	  52.7  &  5.58 $\pm$ 0.10 \\
       & 52.7 &   77.0  &  6.06 $\pm$ 0.76 & $-$18.96±0.13 & 11.6 ± 0.9\\
       & 84.2 &   161.2 &	1.50 $\pm$ 0.27 \\     
\hline
$V$    & 12.2 &   52.7 &  4.52 $\pm$ 0.06  \\
       & 52.7 &   77.0 &  5.36 $\pm$ 0.29  \\
       & 85.9 &  161.2 &  1.38 $\pm$ 0.26  \\
\hline
$r$    & 15.3 &	 32.8 &	 4.05 $\pm$ 0.14  \\
       & 35.4 &  53.4 &  2.65 $\pm$ 0.43 & $-$18.88±0.13 & 11.9 ± 1.6  \\
       & 55.4 &  75.4 &  5.28 $\pm$ 0.65\\
       & 77.0 & 161.2 &  0.89 $\pm$ 0.14  \\
\hline
$i$    & 12.2 &	 54.4 &	 3.41 $\pm$ 0.06  \\
       & 77.0 & 161.2 &  1.62 $\pm$ 0.25  & $-$18.85±0.13 & 12.4 ± 1.7\\
\hline
\end{tabular}
 \end{table}

\begin{table*}
\centering
\caption{Type II SNe comparison sample.} \label{comp_sample}
\renewcommand{\arraystretch}{1.1}
\setlength{\tabcolsep}{3pt}
\scalebox{0.75}{
\begin{tabular}{llllllll}
\hline
SN        & Host   & Explosion  & Redshift & Distance  & E(B$-$V)$_{\rm host}$   & E(B$-$V)$_{\rm MW}$    & References \\
          & galaxy & Epoch (JD) &          &  modulus$^\dagger$ (mag)  & (mag)                   &   (mag) &         \\
\hline
1979C       & NGC 4321                 & 2443970.5 ± 8.0 & 0.00524  & 31.01 ± 0.09           & 0.16 ± 0.05   & 0.0228 ± 0.0003 & 1,2,3,4   \\
1980K       & NGC 6946                 & 2444528.7 ± 5.9 & 0.00016  & 29.44 ± 0.09           & --            & 0.2942 ± 0.0028 & 5, 6, 7   \\
2006Y       & anon                     & 2453766.6 ± 3.4 & 0.0336 & 35.8 & -- & 0.1114 ± 0.0019 & 8,9,10 \\
2006ai      &  ESO 005-G009            & 2453782.1 ± 5.0 & 0.0158 & 35.1 & -- & 0.1080 ± 0.0025 & 8,9,10\\
2014G       & NGC 3448                 & 2456669.8 ± 0.8       & 0.00450  & 31.96 ± 0.14           & 0.268 ± 0.046 & 0.0102 & 11, 12   \\
2016egz     & GALEXASC J000403.88-344851.6 & 2457588.7 ± 2.0   & 0.0232   & 35.11        & --            & 0.0133 ± 0.0003 & 10 \\
2017ahn   & NGC 3318    & 57791.8 ± 0.5   & 0.00925  & 32.70 ± 0.40     & 0.233 ± 0.148 &  0.0667 ± 0.0005 & 13       \\
2018gj      & NGC 6217                 & 2458127.8 ± 1.4 & 0.00454 & 31.46 ± 0.15 & 0.04 ± 0.02 & 0.0375 ± 0.0002 & 14\\
2018hfm     & PGC 1297331              & 2458395.7 ± 5.3       & 0.008    & 32.80 ± 0.60 & 0.26 ± 0.10   & 0.0487 ± 0.0036 & 15 \\
2020jfo     & NGC 4303                 & 2458973.6 ± 1.6 & 0.00522 & 30.81 ± 0.21 & 0.15 ± 0.09 & 0.0194 ± 0.0001 & 16,17,18\\
KSP-SN-2022c & LCRS B234107.3-384735 & 2459773.6 ± 0.01 & 0.041 & 36.4 & -- &   0.0107 ± 0.0003 & 19\\ 
2023ufx     & SDSS J082451.43+211743.3 & 2460223.8 ± 0.5       & 0.0146   & 34.09 ± 0.11 & -- & 0.042 ± 0.0013 & 20\\
2024bch & NGC 3206 & 2460338.0 ± 0.5 & 0.003884 & 31.49 ± 0.45 & 0.029 ± 0.005 & 0.020 ± 0.003 & 21\\
\hline
\end{tabular}}
\tablefoot{
$^\dagger$if redshift dependent, rescaled to H$_0$=67.4 km s$^{-1}$ Mpc$^{-1}$\\
References:  
[1] \cite{Barbon1982B}, [2] \cite{Vaucouleurs1981}, [3] \cite{Balinskaia1980}, [4] \cite{Gall2015}, [5] \cite{Barbon1982A}, [6] \cite{Buta1982}, [7] \cite{Tsvetkov1983}, [8] \cite{Anderson2014}, [9] \cite{Gutirrez2017}, [10] \cite{Hiramatsu2021}, [11] \cite{Bose2016}, [12] \cite{Terreran2016}, [13] \cite{Tartaglia2021}, [14] \cite{Teja2023}, [15] \cite{Zhang2022}, [16] \cite{Sollerman2021}, [17] \cite{Teja2022}, [18] \cite{Ailawadhi2023}, [19] \cite{Jiang2025}, [20] \cite{Ravi2025}, [21] \cite{Andrews2025}.}
 \end{table*}
\begin{table*}
    \caption{Light curve parameters of SN~2021tsz and the comparison sample in $V$-band. \label{sn_detail}}
\renewcommand{\arraystretch}{1.1}
\setlength{\tabcolsep}{3pt}
\scriptsize
     \begin{tabular}{lllllllllllll}
\hline
SN & a$_0$ (mag)  & w$_0$ (d) &  t$_{\rm PT}$ (d) & s$_1$ (mag/100d) & t$_{\rm range}$ & s$_2$ (mag/100d) & t$_{\rm range}$ & s$_3$ (mag/100d) & t$_{\rm range}$     &   M$^{\rm max}$ & M$_{\rm Ni}^\ddagger$  (M$_\odot$)\\
 \hline
1979C & -- & -- & -- & 2.55 ± 0.02 & 16.7-44.6 & --          & --         & 3.02 ± 0.01 & 53.6-118.7  & $-$19.48 ± 0.19$^\dagger$ & --\\
1980K & -- & -- & -- & 3.98 ± 0.47 & 15.5-73.6 & --          & --         & 1.20 ± 0.35 & 112.6-285.8 & $-$18.90 ± 0.10$^\dagger$ & 0.0365 ± 0.0127\\
2006Y   & 1.8$\pm$0.2 & 4.2$^{+0.7}_{-0.6}$ & 60.1$^{+3.6}_{-3.5}$ & 6.1$\pm$0.1 & 8.6-23.2 & 0.41$\pm$0.37 & 31.9-41.6 & -- & -- & $-$18.09 $\pm$ 0.01$^\dagger$ & --\\ 
2006ai  & 1.54$\pm$0.08 & 3.7$\pm$0.4 & 71.1$\pm$0.3 & 4.7$\pm$0.1 & 13.3-22.3 & 2.12$\pm$0.04 & 27.1-62.4 & 1.8$\pm$0.3 & 78.1-107.6 & $-$18.19 $\pm$ 0.01$^\dagger$ & 0.062$\pm$0.002\\
2014G & 2.08$\pm$0.55 & 4.57 $\pm$ 1.57 & 85.9$\pm$0.7 & 2.87 ± 0.05 & 12.5-77.5 & -- & -- & 1.68 ± 0.18 & 95.2-289.4 & $-$18.53$\pm$0.25 & 0.07258 ± 0.01297\\
2016egz &  2.15$\pm$0.30 & 3.3$^{+1.2}_{-0.8}$ & 71.9$^{+2.1}_{-2.2}$ & 6.65 $\pm$ 0.35 & 7.4-17.7 & 1.44 $\pm$ 0.09 & 21.4-64.5 & 0.64 $\pm$ 0.19 & 93.1-174.8 & $-$18.46 $\pm$ 0.03$^\dagger$  &  0.090 $\pm$ 0.005 \\
2017ahn & 2.60$\pm$0.27 & 9.49$\pm$1.23 & 54.2$\pm$0.9 & 4.11$\pm$0.06 & 7.4-52.8 & -- & -- & 1.96$\pm$0.05 & 72.3-135.6 & $-$18.40 $\pm$ 0.50 & 0.05059 ± 0.01574 \\
2018gj & 1.48$^{+0.07}_{-0.06}$ & 3.2$\pm$0.4 & 78.1$\pm$1.5 & 2.24$\pm$0.08 & 4.7-17.7 & 1.35$\pm$0.03 & 23.7-56.7 & 1.31$\pm$0.01 & 92.7-295.3 & $-$17.0$\pm$0.1$^\dagger$ & 0.026$\pm$0.007\\
2018hfm & 2.46$\pm$0.26 & 3.01$\pm$0.67 & 57.4$\pm$0.7 & 4.40 ± 0.21 & 5.4-41.3 & -- & -- & -- & -- & $-$18.72 $\pm$ 0.69$^\dagger$ & $<$0.015\\
2020jfo & 1.80±0.16 & 4.82±0.92 & 67.6±1.8 & 1.19 ± 0.06 & 13.7-46.4  & -- & -- & 1.37 ± 0.03 & 93.6-345.8  & $-$16.87$\pm$0.33 & 0.03 $\pm$ 0.01\\
2021tsz & 0.97 $\pm$ 0.08 (r) & 4.6$^{+0.3}_{-0.6}$ (r) & 62.5$^{+1.3}_{-1.2}$ (r) & 4.52 $\pm$ 0.06 & 12.2-52.7 & -- & -- & 0.89$\pm$0.14 & 77.0-161.2 & $-$18.84$\pm$0.04 (r) &  0.08$\pm$0.01  \\
2023ufx & 1.16$\pm$0.08 & 4.6$\pm$0.6 & 46.5$\pm$0.5 & 6.96$\pm$0.07 & 7.4-21.6 & 1.06$\pm$0.10 & 23.3-37.9 & 1.72$\pm$0.03 & 63.5-212.1 & $-$18.42$\pm$0.08  & 0.137$^{+0.019}_{-0.017}$\\
2024bch & 1.50$^{+0.04}_{-0.05}$ & 3.6$\pm$0.3 & 78.0$\pm$0.5 & 3.3$\pm$0.3 & 23.4-45.6 & 1.8$\pm$0.3 & 48.4-66.7 & 1.45$\pm$0.03 & 89.4-153.2 & $-$17.8$\pm$0.45 & 0.055$\pm$0.002\\
\hline
\end{tabular}
\tablefoot{\scriptsize $^\dagger$lower limit; peak not observed.
\scriptsize $^\ddagger$obtained from \cite{Rodriguez2021} or the references provided in Table~\ref{comp_sample}.}
\end{table*}

\end{appendix}
\end{document}